\documentclass[twocolumn,times,twocolappendix]{aastex631}
\usepackage{amsmath,bm}
\usepackage{xcolor}
\usepackage[english]{babel}
\usepackage[latin1]{inputenc} 
\usepackage{enumitem}
\usepackage{float}
\usepackage[title]{appendix}
\usepackage{natbib}

\newcommand{\mpchi}{\,h^{-1}{\rm {Mpc}}}

\newcommand{\kms}{{\rm km\,s^{-1}}}
\newcommand{\msun}{M_{\sun}}

\newcommand{\hi}{H{~\sc i}}

\shorttitle{Cold Gas Depletion from the AGN Feedback}
\shortauthors{W. Ma et al.}

\begin{document}
	
	\title{Effects of Active Galactic Nucleus Feedback on Cold Gas Depletion and Quenching of Central Galaxies}
	
	\author[0000-0003-4978-5569]{Wenlin Ma}
	\affiliation{Shanghai Astronomical Observatory, Chinese Academy of Sciences, Shanghai 200030, China}
	\affiliation{University of Chinese Academy of Sciences, Beijing 100049, China}
	\email{mawenlin@shao.ac.cn; guohong@shao.ac.cn}
	
	\author[0000-0002-8604-2556]{Kexin Liu}
	\affiliation{Shanghai Astronomical Observatory, Chinese Academy of Sciences, Shanghai 200030, China}
	\affiliation{University of Chinese Academy of Sciences, Beijing 100049, China}
	
	\author[0000-0003-4936-8247]{Hong Guo}
	\affiliation{Shanghai Astronomical Observatory, Chinese Academy of Sciences, Shanghai 200030, China}
	\affiliation{Corresponding Author}
	
	\author[0000-0002-2113-4863]{Weiguang Cui}
	\affiliation{Institute for Astronomy, University of Edinburgh, Edinburgh, UK}
	\affiliation{Departamento de F\'isica Te\'{o}rica, M\'{o}dulo 15, Facultad de Ciencias, Universidad Aut\'{o}noma de Madrid, 28049 Madrid, Spain}
	
	\author[0000-0002-5434-4904]{Michael G. Jones}
	\affiliation{Steward Observatory, University of Arizona, 933 North Cherry Avenue, Rm. N204, Tucson, AZ 85721-0065, USA}
	
	\author[0000-0002-6593-8820]{Jing Wang}
	\affiliation{Kavli Institute for Astronomy and Astrophysics, Peking University, Beijing 100871, China}
	
	\author{Le Zhang}
	\affiliation{Shanghai Astronomical Observatory, Chinese Academy of Sciences, Shanghai 200030, China}
	\affiliation{University of Chinese Academy of Sciences, Beijing 100049, China}
	
	\author[0000-0003-2842-9434]{Romeel Dav\'e}
	\affiliation{Institute for Astronomy, University of Edinburgh, Edinburgh, UK}
	\affiliation{University of the Western Cape, Bellville, Cape Town 7535, South Africa}
	\affiliation{South African Astronomical Observatories, Observatory, Cape Town 7925, South Africa}
	
	\begin{abstract}
		We investigate the influence of active galactic nucleus (AGN) feedback on the galaxy cold gas content and its connection to galaxy quenching in three hydrodynamical simulations of Illustris, IllustrisTNG and SIMBA. By comparing to the observed atomic and molecular neutral hydrogen measurements for central galaxies, we find that Illustris over-predicts the cold gas masses in star-forming galaxies and significantly under-predicts them for quenched galaxies. IllustrisTNG performs better in this comparison than Illustris, but quenched galaxies retain too much cold gas compared with observations. SIMBA shows good agreement with observations, by depleting the global cold gas reservoir for quenched galaxies. We find that the discrepancies in IllustrisTNG are caused by its weak kinetic AGN feedback that only redistributes the cold gas from the inner disks to the outer regions and reduces the inner cold gas densities. It agrees with observations much better when only the cold gas within the stellar disk is considered to infer the star formation rates. From dependences of cold gas reservoir on the black hole mass and Eddington ratio, we find that the cumulative energy release during the black hole growth is the dominant reason for the cold gas depletion and thus the galaxy quenching. We further measure the central stellar surface  density within 1~kpc ($\Sigma_1$) for the high-resolution run of IllustrisTNG and find a tight correlation between $\Sigma_1$ and black hole mass. It suggests that the observed decreasing trend of cold gas mass with $\Sigma_1$ is also a reflection of the black hole growth.
	\end{abstract}

	\section{Introduction}
	How massive galaxies quench their star formation is one of the key questions in the study of galaxy evolution. Various physical mechanisms have been proposed to understand the quenching process \citep[see e.g.,][]{Birnboim2003,Dekel2006,Bower2006,Croton2006,Martig2009,Ishibashi2012,Zolotov2015}. Among them, the feedback from active galactic nuclei (AGNs) is thought to be one of the most effective channels to shut down  star formation \citep[see ][and references therein]{Heckman2014}. In fact, studies using hydrodynamical simulations and semi-analytic models have shown that the observed galaxy properties, such as the stellar mass function, quenched fraction and morphology, can be reasonably reproduced only when AGN feedback is included in the galaxy formation models \citep[see e.g.,][] {Matteo2005,McCarthy2011,Dubois2016, Beckmann2017,Kaviraj2017,Donnari2021}. 
	
	Galaxy star formation is fueled by the cold gas supply, as indicated in the empirical relation between the star formation rate (SFR) surface density and the cold gas surface density, well known as the Kennicutt-Schmidt law \citep{Schmidt1959,Kennicutt1998}. It is generally agreed that the atomic neutral hydrogen (\hi) gas needs to be converted to the molecular hydrogen (H$_2$) to form stars \citep{Wong2002,Kennicutt2007,Bigiel2008,Leroy2008}. Thus, star formation quenching could be due to the depletion of the \hi\ gas reservoir, the prevention of conversion from \hi\ to H$_2$ or the low star formation efficiency (${\rm SFR}/M_{\rm H_2}$) \citep{Man2018} . The influence of AGN feedback in all these processes is of special importance for the galaxy formation models. 
	
	With the implementation of reasonable subgrid physics of star formation and feedback mechanisms, the modern hydrodynamical simulations generally agree well with the observed properties of galaxy stellar components \citep[e.g.,][]{Vogelsberger2014,Schaye2015,Pillepich2018,Dave2019}. However, it is becoming more challenging for them to reproduce the observations of neutral hydrogen gas \citep{Crain2015,Guo2017,Guo2020,Diemer2019,Dave2020}, including the cold gas mass functions \citep[e.g.,][]{Keres2003,Zwaan2005,Martin2010,Jones2018,Fletcher2021} and the cold gas scaling relations \citep[see][for a review]{Saintonge2022}.
	
	The cold gas content in galaxies can also provide important clues to the causes of galaxy quenching \citep[e.g.,][]{Appleby2020,Dave2020,Davies2020,Piotrowska2022,Ward2022}. However, a key limitation of the current cold gas observations is the lack of large and homogeneous catalogs that fairly sample the star-forming and quenched galaxies. The less cold gas content in the quenched galaxies makes their \hi\ or H$_2$ signals typically below the detection limits of the relevant surveys. The spectra stacking technique can be adopted to bypass this limitation to obtain the average cold gas masses for different populations \citep[see e.g.,][]{Fabello2011,Saintonge2016,Ellison2019,Guo2020}. 
	
	By spectra stacking the \hi\ signals of star-forming and quenched galaxies, \cite{Guo2021} (hereafter G21) found that star formation in the local universe is directly regulated by the available \hi\ gas reservoir, extending the previous results based on much smaller samples \citep{Saintonge2016}. Therefore, measuring the \hi\ and H$_2$ gas in AGN and non-AGN host galaxies is helpful for understanding the influence of AGN feedback. However, direct observational probes of the gas content reveal no strong differences between AGN and non-AGN hosts, for \hi\ \citep{Fabello2011,Gereb2015,Ellison2019} and H$_2$ gas \citep{Saintonge2017,Shangguan2020}. Indirect probes of the total cold gas mass based on the gas-to-dust ratio have reported similar or even higher gas masses of AGN hosts \citep{Vito2014,Shangguan2018,Shangguan2019}. It has been suggested that AGN feedback might only directly affect the cold gas in the central regions of galaxies, and is ineffective at clearing out gas at larger radii and thus has little effect on the total gas mass \citep{Fluetsch2019,Ellison2021}. While the short timescale of AGN activity may play a role in the observed gas properties of AGN and non-AGN hosts, the self-regulating black hole growth \citep{Heckman2014} would also suggest that galaxies with higher cold gas content tend to have stronger AGN feedback, as shown in some recent works \citep{Koss2021,Guo2022}.
	
	In order to better understand the role of AGN in star formation quenching, in this paper, we make use of three state-of-the-art hydrodynamical simulations, Illustris \citep{Vogelsberger2014}, IllustrisTNG  \citep{Marinacci2018,Naiman2018,Nelson2018,Pillepich2018,Springel2018} and SIMBA \citep{Dave2019}. By comparing their simulated \hi\ gas masses in star-forming and quenched galaxies with the observed values of G21, we are able to investigate the potential observational effects of AGN feedback in different galaxy formation models. Since the gas content of satellite galaxies suffers from severe environmental effects, like ram-pressure and tidal stripping \citep{Brown2017,Cortese2021,Wang2021,Wang2022}, we focus on studies of central galaxies (i.e. locating at the centers of their host dark matter halos) as in G21. 
	
	The organization of this paper is as follows. In Section~\ref{sec:data}, we introduce the simulations and observational data we used. We show the results in Section~\ref{sec:results}. Discussions and conclusions are presented in Section~\ref{sec:discussion} and Section~\ref{sec:conclusion}, respectively. Throughout the paper, the stellar and \hi\ masses are all in units of $\msun$. We assume a flat $\Lambda$CDM cosmology of $\Omega_m=0.307$, $\Omega_\Lambda=0.693$, and $h=0.7$, consistent with the \hi\ observations. 
	
	\section{Data and Method}\label{sec:data}
	
	\subsection{Observational Measurements}\label{sec:obs_g21}
	We adopt the observational measurements of average \hi\ masses for star-forming galaxies (hereafter SFGs) and quenched galaxies (hereafter QGs) from G21 for central galaxies only. They were obtained from stacking the \hi\ spectra \citep{Guo2020} in the overlap regions between the \hi\ data of Arecibo Fast Legacy ALFA Survey \citep[ALFALFA;][]{Giovanelli2005,Haynes2011,Haynes2018} and the optical main galaxy sample of the Sloan Digital Sky Survey \citep[SDSS;][]{York2000} DR7 \citep{Abazajian2009}. The central galaxies are identified using the galaxy group catalog of \cite{Lim2017}. Due to the \hi\ flux limit of the ALFALFA survey, their measurements are only in the redshift range of $0.0025<z<0.06$, which can be directly compared with the simulation outputs at $z\sim0$. The accuracy of such an \hi\ stacking method has been verified using mock galaxy catalogs \citep{Chauhan2021}.
	
    The \hi\ mass measurements in G21 are in fact $\log\langle M_{\rm HI}\rangle$, as the average is obtained by co-adding the \hi\ fluxes of stacked galaxies. We refer readers to G21 for the details of \hi\ spectra stacking. For fair comparisons, we will also measure the same quantity in the following simulations. The galaxy stellar mass and star formation rate in G21 are adopted from the GSWLC-2 catalog \citep{Salim2018} using the UV/optical spectral energy distribution (SED) fitting to the photometry of SDSS galaxies.
	
	\subsection{Illustris simulation}
	In this paper, we use the highest-resolution simulation of the Illustris suite, Illustris-1 \citep{Nelson2015}, which was run using the {\scriptsize AREPO} code \citep{Springel2010}, with a volume of $75^3{\rm Mpc/h}^3$. The dark matter and baryon mass resolutions are $6.26\times10^6\msun$ and $1.26\times10^6\msun$, respectively \citep{Vogelsberger2014}. The AGN feedback model in Illustris is comprised of three different modes, i.e. quasar, radio and radiative modes \citep{Sijacki2015}. In the high-accretion quasar mode, a small fraction of the AGN bolometric luminosity is thermally and isotropically coupled to the surrounding gas with an efficiency of 0.05, which would lead to effective energy-driven outflows. 
		
	The most important feature of the Illustris AGN model is in the low-accretion radio mode, where hot bubbles are randomly injected into the circumgalactic medium (CGM) within a sphere around each black hole \citep{Sijacki2007}. The sphere radius is twice the bubble radius and the injected thermal energy is determined by,
    \begin{equation}
		E_{\rm bub}=\epsilon_{\rm m}\epsilon_{\rm r}c^2 \delta M_{\rm BH},
    \end{equation}
	where $\epsilon_{\rm m}=0.35$ is the efficiency of mechanical heating by bubbles, the radiative efficiency $\epsilon_{\rm r}$ is set as 0.2 \citep{Sijacki2015}, $\delta M_{\rm BH}$ is the increased black hole mass, and $c$ is the speed of light. The bubble radius is calculated from both $E_{\rm bub}$ and the density of surrounding CGM.
	
	The transition from quasar to radio mode in Illustris happens when the Eddington ratio $f_{\rm edd}$ drops below a threshold of 0.05. The Eddington ratio is defined as,
	\begin{equation}
		f_{\rm edd} = \frac{\dot{M}_{\rm BH}}{\dot{M}_{\rm Edd}},
	\end{equation}
	with $\dot{M}_{\rm Edd}$ being the Eddington accretion rate. The radiative AGN feedback in Illustris is added by modifying the net gas cooling rate in the vicinity of the black hole particle. 
	
	Though Illustris can explain observations well in many aspects, it is found that its gas fractions of galaxy groups and clusters are too low \citep{Genel2014}. This may be ascribed to the excessive gas removal by the thermal bubbles injected in the radio mode, as also found in this work.
	
	\subsection{IllustrisTNG simulation}\label{subsec: TNG simulation}
	The IllustrisTNG (hereafter TNG) simulation \citep{Nelson2019} is improved upon Illustris in both the numerical techniques of the {\scriptsize AREPO} code \citep{Pakmor2016} and the galaxy formation model \citep{Weinberger2017,Pillepich2018}. The most significant difference is that the radio bubble feedback in Illustris is replaced with the kinetic wind feedback in TNG. As in Illustris, the thermal energy is injected in a small volume around each black hole for the high-accretion state in the thermal mode. But in the low-accretion state, once the black holes accumulate enough energy above a certain threshold, the kinetic energy will be released as a momentum boost to the gas cells in the feedback region. The feedback energy is formulated as
	\begin{equation}
		\Delta\dot{E}_{\rm low}=\epsilon_{\rm f,kin}\dot{M}_{\rm BH}c^2
	\end{equation}
	where the coupling efficiency $\epsilon_{\rm f,kin}$ is proportional to the surrounding gas density and has a maximum of $0.2$ \citep[more details are presented in][]{Weinberger2017}.
	
	The threshold $\chi$ of $f_{\rm edd}$ for the transition from thermal mode to kinetic mode is determined by,
	\begin{equation}
		\chi = \min[0.002(M_{\rm BH}/10^8\msun)^2,0.1].
	\end{equation}
	As shown in \cite{Terrazas2020}, the AGN feedback will be gradually dominated by the kinetic mode when $M_{\rm BH}$ exceeds $10^{8.2}\msun$.
	
	In this paper, we focus on the TNG100-1 simulation, with a box size of $75\mpchi$ on a side. The dark matter particle and gas cell masses are slightly larger than those in Illustris due to the use of Planck2015 cosmology instead of WMAP-9. For both Illustris and TNG, the stellar mass for each galaxy is defined as the total mass of stellar particles within twice the stellar half mass radius (i.e. $2R_{\rm half}$) \citep{Nelson2019}. To ensure robust measurements of galaxy properties, we have required the lower limit of galaxy stellar mass to be $10^9\msun$ for our final sample. The final galaxy sample sizes for Illustris and TNG100-1 are 20742 and 10942, respectively.
	
	Star formation in Illustris and TNG is regulated by a modified two-phase interstellar medium (ISM) model of \cite{Springel2003}, where stochastic star formation occurs in gas cells exceeding a given density threshold of $n_{\rm H}=0.106\,{\rm cm}^{-3}$. To obtain the \hi\ gas mass in both Illustris and TNG, we use the post-processing framework presented in \cite{Diemer2018}, where five different models have been applied for the atomic-to-molecular transition. The \hi\ and H$_2$ masses are measured within the whole subhalo containing the corresponding galaxy. We explore the effects of different atomic-to-molecular transition models in Appendix~\ref{app}. The results of different \hi\ models are quite consistent with each other, as also confirmed in \cite{Diemer2019}. While there are slightly larger differences for the H$_2$ measurements, our conclusions are not affected by the model variations.
	
	We only use the output of `K13' model with the projected quantities \citep{Krumholz2013} in \cite{Diemer2018}. \cite{Krumholz2013} took the varying interstellar radiation field and cold gas density within interstellar medium (ISM) into consideration, as they will affect the molecular fraction of the neutral hydrogen ($f_{\rm H_2}$) and SFR. In this prescription, $f_{\rm H_2}$ is determined from the UV background ($\rm{U_{MW}}$) and cold phase column density of ISM ($\rm{n_{CNM}}$), as follows
	\begin{equation}
		f_{\rm H_2} =\begin{cases}
                        1-3s/(4+s) & \text{if $s<2$} \\
                        0          & \text{if $s \geq 2$}
                       \end{cases},
                   \label{eq:fh2}
	\end{equation}
	where 
	\begin{equation}
		s\equiv \rm{ \frac{\ln{(1+0.6\chi+0.01\chi^2)}}{0.6\tau_c}} ,  \label{eq:s}
	\end{equation}
	and
	\begin{equation}
		\chi \equiv \rm{7.2U_{MW}( \frac{n_{CNM}}{10cm^{-3}})^{-1}} , \label{eq:chi}
	\end{equation}
	with $\tau_c$ being the optical depth of a cloud. Since the UV field is estimated from the input SFR, the post-processed cold gas mass in TNG will not be self-consistent at different redshift outputs. We will further discuss this at Section~\ref{sec:discussion}. 
	We refer the readers to \cite{Diemer2018} for more details. 
	
	To investigate the small-scale structural differences between star-forming and quenched populations, we also use the TNG50-1 simulation \citep{Nelson2019b,Pillepich2019}, the highest mass resolution of the TNG suite, for calculating the central stellar surface density within 1~kpc (referred to as the $\Sigma_1$ parameter) in Section~\ref{subsec:sigma1}. Its box size is $35\mpchi$ on a side and the mass resolution is about $8.5\times10^4\msun$, 16.5 times higher than that of TNG100-1, allowing to resolve internal details of galaxies. We will further make use of $M_\ast$, SFR, $M_{\rm BH}$, \hi\ and H$_2$ masses (as well as their density profiles) from the public release of TNG50-1 in Section~\ref{subsec:sigma1}. These parameters are measured in the same way as in TNG100-1. The final sample size for TNG50-1 is 1618.
	
	\begin{figure*}
		\centering
		\includegraphics[width=1\textwidth]{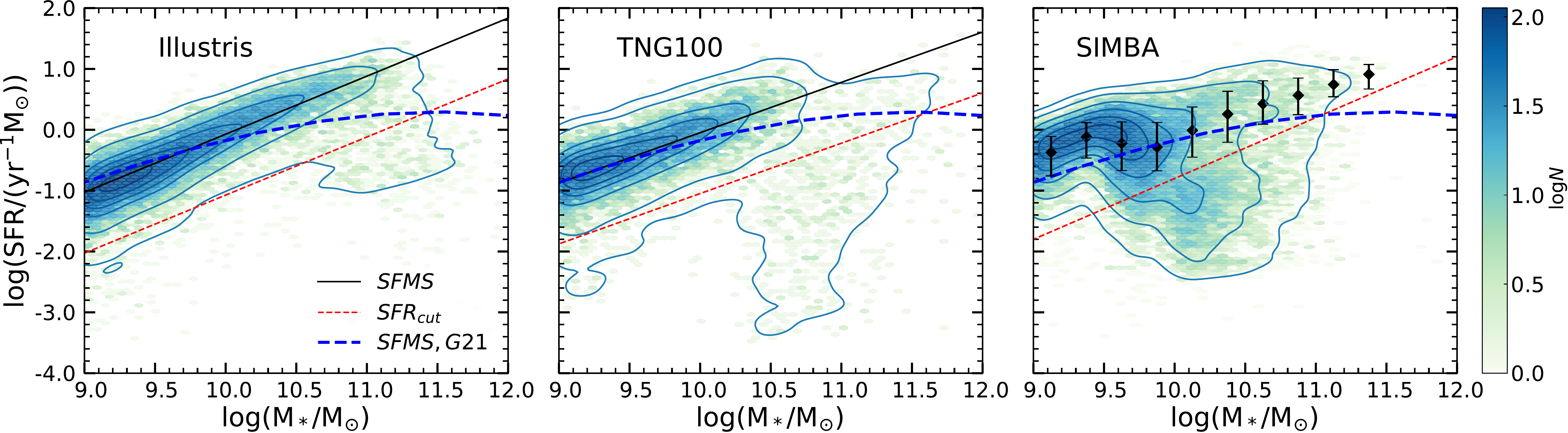}
		\caption{Distribution of central galaxies as a function of SFR and $M_{\ast}$. Each panel from left to right is Illustris, TNG and SIMBA, respectively. The bins are color-coded by the logarithmic number of galaxies. Black lines are the SFMS for Illustris and TNG, while black dots are shown in SIMBA. Red dotted lines are the separation of star forming and quenched populations for all simulations (Eqs.~\ref{eq:cut_ill}--\ref{eq:cut_simba}). The observed SFMS of G21 is shown as the blue dashed line in each panel.}
		\label{fig:SFMS}
	\end{figure*}
	\subsection{SIMBA simulation}\label{subsec: SIMBA simulation}
	The SIMBA simulations \citep{Dave2019} are based on the {\scriptsize MUFASA} suite of cosmological hydrodynamic simulations \citep{Dave2016}, running with the {\scriptsize GIZMO} code \citep[][]{Hopkins2015}. We adopt the simulation run of m100n1024, with a box size of $100\mpchi$ on a side. The dark matter and the initial baryon  mass resolutions are $9.6\times10^7\msun$ and $1.82\times10^7\msun$, respectively. These mass resolutions are about 13 times poorer than TNG100-1. The galaxies in SIMBA is identified on the fly using a 6D friends-of-friends algorithm with a linking length $0.0056$ times the mean particle separation. Each gas particle in a halo is assigned to the galaxy that has the largest value of $M_{\rm baryon}/R^2$, with $M_{\rm baryon}$ being the total baryonic mass and $R$ the distance to galaxy center.  The final sample includes 23759 galaxies with $M_\ast>10^9\msun$, which corresponds to about 55 star particles.
	
	Different from Illustris and TNG, where the black hole accretion is parametrized in terms of Eddington limited Bondi-Hoyle-Lyttleton accretion \citep{Matteo2005}, SIMBA uses a two-mode black hole accretion model, i.e. torque-limited accretion for cold gas ($T<10^5K$) and Bondi accretion for hot gas ($T>10^5K$). In the former mode, the gas inflow rate $\dot{M}_{\rm Torque}$ is driven by disk gravitational instabilities following \cite{Hopkins2011} \citep[see also][]{Angles2013,Angles2015,Angles2017} and multiplied by an efficiency of 0.1 to match the local $M_{\rm BH}$--$M_\ast$ relation, while in the latter mode, the standard Bondi accretion rate is suppressed by the same efficiency of $0.1$ as in $\dot{M}_{\rm Torque}$.
	
	The black hole feedback mechanism in SIMBA consists of a kinetic feedback and an X-ray energy feedback. The kinetic subgrid model in SIMBA is designed to mimic the observed two-mode AGN feedback, with the high-accretion and low-accretion modes divided by the Eddington ratio of $f_{\rm edd}=0.2$. In the high-accretion mode, a $M_{\rm BH}$-dependent outflow velocity is applied to gas surrounding the black hole, with a typical velocity less than $1000\,\kms$, mimicking radiative AGN winds. In the low-accretion mode (jet mode) with $f_{\rm edd}<0.2$ and $M_{\rm BH}>10^{7.5}\msun$, the outflow velocity significantly increases with decreasing $f_{\rm edd}$, with the full-speed jets ($\sim 8000\,\kms$) achieved at $f_{\rm edd}<0.02$. When the full-speed jet is activated and the gas fraction $M_{\rm gas}/M_\ast$ is lower than 0.2, the X-ray feedback will be introduced by injecting additional energy into the surrounding gas. We refer the readers to \cite{Dave2019} for more details. 
	
	The star formation in SIMBA is directly modeled with the Kennicutt-Schmidt law by calculating the molecular gas fraction of the total gas, $f^\prime_{\rm H_2}$, following the subgrid model of \cite{Krumholz2011}, similarly as in Equations~(\ref{eq:fh2})--(\ref{eq:chi}) \citep{Dave2020}. We emphasize that $f^\prime_{\rm H_2}$ here is the molecular fraction for any given gas (including helium and metals), rather than just neutral hydrogen as in TNG. But the parameter $\chi$ of Equation~(\ref{eq:s}) in SIMBA is a function of local metallicity in the gas cell, rather than estimated post-processingly from SFR as in Illustris and TNG.  Moreover, the SFR for a given gas element in SIMBA is calculated as 
		\begin{equation}
        {\rm SFR}=0.02f^\prime_{\rm H_2}\rho/t_{\rm dyn},	\label{eq:sfr_simba}			
		\end{equation}
	where $\rho$ is the gas density and $t_{\rm dyn}=1/\sqrt{G\rho}$ is the local dynamical time. The \hi\ fraction of a given gas element is computed using the prescription of \cite{Rahmati2013}, accounting for the self-shielding effect. Adding the \hi\ and H$_2$ fractions gives the total neutral gas fraction. In this way, the galaxy cold gas mass can be self-consistently computed on the fly during the simulation run of SIMBA \citep{Dave2020}. 
	
	\begin{figure*}
		\centering
		\includegraphics[width=1\textwidth]{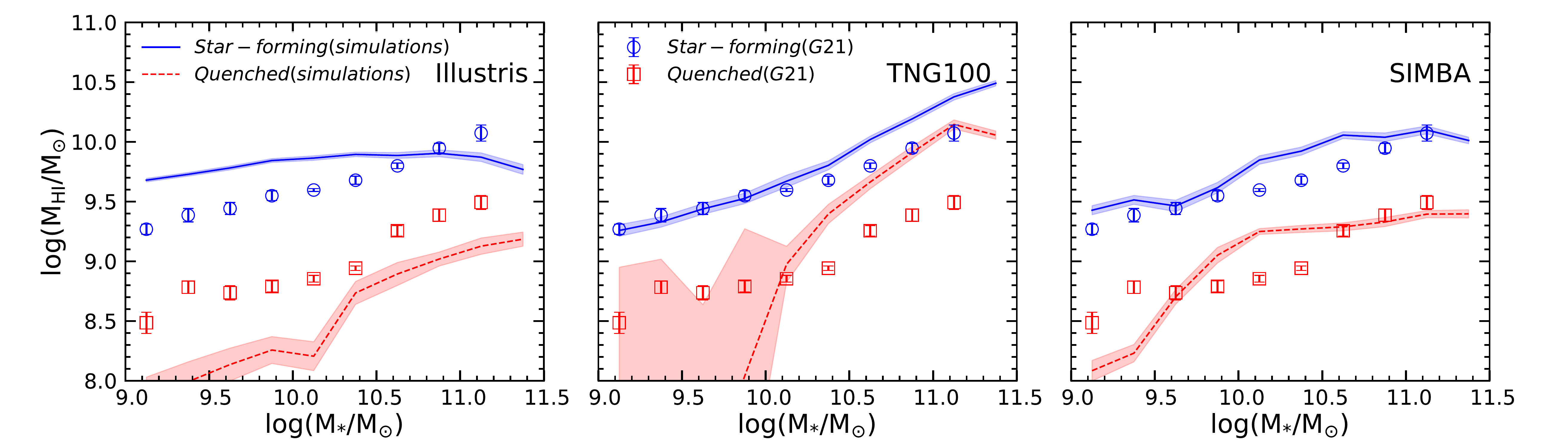}
		\caption{Average \hi\ mass of central galaxies as a function of stellar mass. From left to right, the results are shown for SFGs (blue) and QGs (red) in Illustris, TNG100 and SIMBA, respectively. We show the simulation predictions as solid (SFGs) and dotted (QGs) lines with the shaded regions representing the 1$\sigma$ error distributions using bootstrapping method, while the observational measurements of \cite{Guo2021} are displayed with circles for SFGs and squares for QGs.} 
		\label{fig:hi-mstar}
	\end{figure*}

	\subsection{Star Formation Main Sequence}\label{subsec:SFMS}
	To compare with the observational \hi\ mass measurements for SFGs and QGs, we similarly divide the galaxy samples in the simulations into two populations. For each simulation, we only select central galaxies with $M_\ast>10^9\msun$ at $z=0$, which are fully resolved for the different simulation resolutions. We show the distributions of galaxy samples in the SFR--$M_\ast$ plane in Figure~\ref{fig:SFMS} for Illustris (left panel), TNG (middle panel) and SIMBA (right panel), respectively. As is common practice, we define the star formation main sequence (SFMS) by fitting the number density peaks for galaxies with specific SFR (sSFR) larger than $10^{-11}{\rm yr}^{-1}$ using a simple power law for Illustris and TNG. The best-fitting relations for the SFMS are shown as the solid black lines in the left and middle panels. Since the SFMS for SIMBA is apparently curved for $M_\ast>10^{9.5}\msun$, we only use the median values of SFR in each $M_\ast$ bin for galaxies with ${\rm sSFR}>10^{-10.8}{\rm yr}^{-1}$ as the SFMS and use interpolation for the relevant calculations, rather than fitting it with a functional form. The best-fitting SFMS relations for Illustris and TNG are,
    \begin{eqnarray}
		\log({\rm SFR}_{\rm MS,Illustris}/{\rm yr}^{-1}\msun) &=& 0.95\log M_\ast - 8.62, \label{eq:ms_ill}\\
		\log({\rm SFR}_{\rm MS,TNG}/{\rm yr}^{-1}\msun) &=& 0.83\log M_\ast-8.32. \label{eq:ms_tng}
	\end{eqnarray}

    Our fitting parameters for SFMS are consistent with previous studies \citep{Weinberger2018,Donnari2019,Donnari2021,Hahn2019}. For example, \cite{Hahn2019} used $\log{\rm SFR}_{\rm MS}=1.01\log M_\ast-10.02$ for galaxies in Illustris and  \cite{Donnari2019} adopted a similar definition of $\log{\rm SFR}_{\rm MS}=0.8\log M_\ast-8.15$ for TNG. 	
	
    The observational definition of SFMS in G21 (shown as dashed blue lines) is to fit the average SFR for SFGs with a third-order polynomial as,
    \begin{eqnarray}
		\log({\rm SFR}_{\rm MS,G21}/{\rm yr}^{-1}\msun) &=& -2.61\log M_\ast + 0.46(\log M_\ast)^2\nonumber \\ 
		&&- 0.02(\log M_\ast)^3, \label{eq:ms_G21}
	\end{eqnarray}
    where the slope of SFMS is gradually becoming flatter for more massive galaxies. Their cut to separate SFGs and QGs is
	\begin{eqnarray}
		\log({\rm SFR}_{\rm cut,G21}/{\rm yr}^{-1}\msun) &=& 0.65\log M_\ast - 7.25 . \label{eq:cut_G21}
	\end{eqnarray}
	
	To be consistent with literature, for Illustris and TNG, the cuts between SFGs and QGs are simply 1~dex below the SFMS, 
	\begin{eqnarray}
		\log({\rm SFR}_{\rm cut,Illustris}/{\rm yr}^{-1}\msun) &=& 0.95\log M_\ast - 9.62 \label{eq:cut_ill}\\
		\log({\rm SFR}_{\rm cut,TNG}/{\rm yr}^{-1}\msun) &=& 0.83\log M_\ast-9.32. \label{eq:cut_tng}	
	\end{eqnarray}
    For SIMBA, we follow the definition of \cite{Dave2019} as 
	\begin{equation}
	    \log({\rm SFR}_{\rm cut,SIMBA}/{\rm yr}^{-1}\msun) = \log M_\ast-10.8 . \label{eq:cut_simba}
	\end{equation}
	
	The conclusions of our study are not significantly affected by the definitions of the cuts, as they are all around 3$\sigma$ below the SFMS in each simulation. We note that all galaxies below these cuts are regarded as QGs, including those without SFR measurements (i.e., SFR$=0$) due to the simulation resolution limits. 
	
	The distributions of central galaxies in the SFR-$M_\ast$ plane for the three simulations are quite different from each other. But QGs are well separated from SFGs with the cuts applied. We expect that the low-mass QGs of $M_\ast<10^{9.5}\msun$ are more likely the low-SFR tail of SFGs. We will focus on the massive galaxies of $M_\ast>10^{9.5}\msun$ in this study.
	
	\section{Results}\label{sec:results}
	\subsection{H{~\scriptsize I}--Stellar Mass Relation}
	\begin{figure*}
		\centering
		\includegraphics[width=0.8\textwidth]{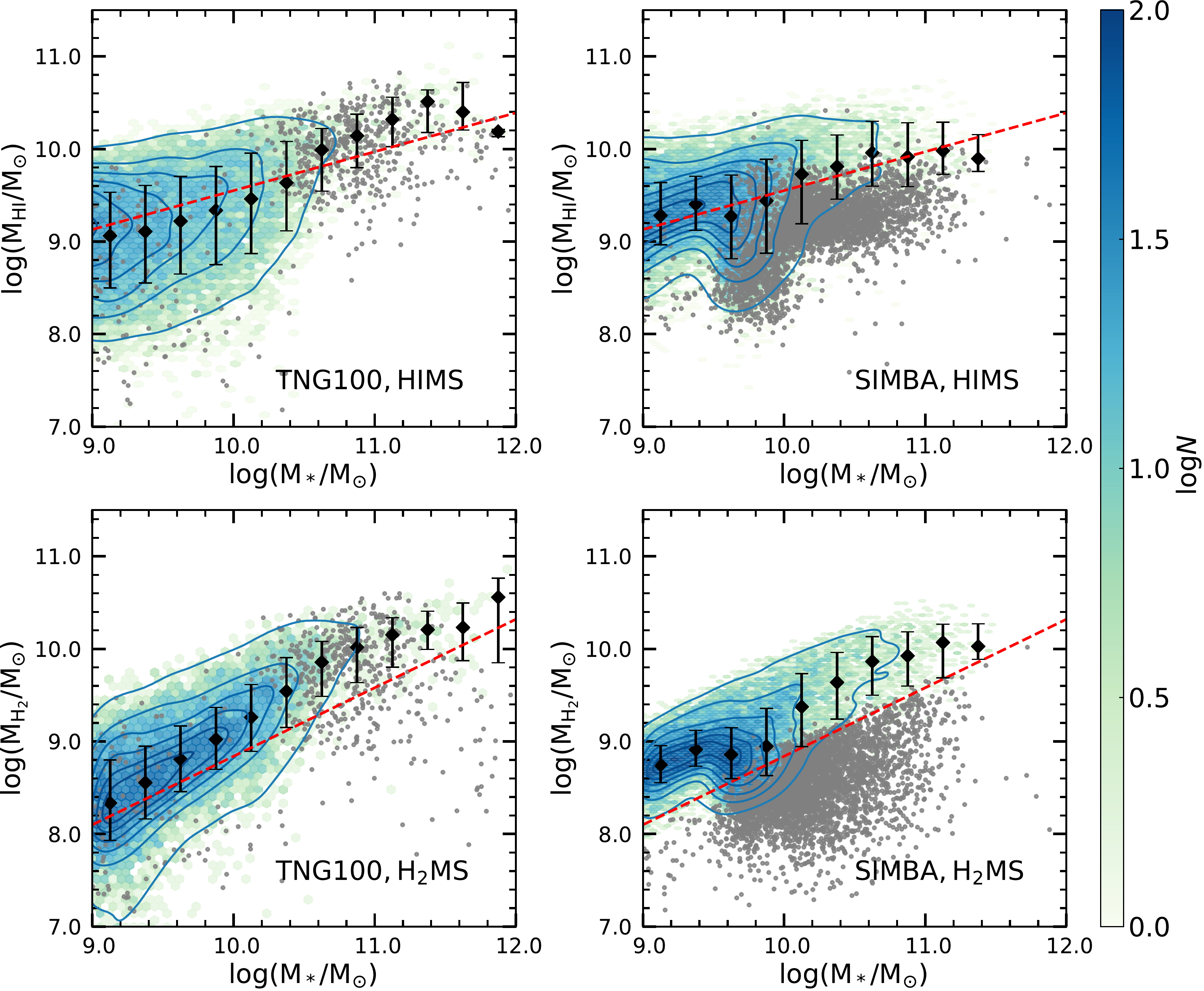}
		\caption{Distribution of galaxies in the planes of $M_{\rm HI}$--$M_\ast$ (top panels) and $M_{\rm H_2}$--$M_\ast$ (bottom panels) for TNG (left) and SIMBA (right), respectively. The \hi\ and H$_2$ main sequences are defined as the corresponding median values for SFGs in different $M_{\ast}$ bins, shown as the black points. Distributions of SFGs and QGs are represented by the contour and gray dots in each panel. The probability distribution of SFGs is represented by the logarithmic color scales. The red dotted lines are HIMS and H$_2$MS defined in G21 and \cite{Janowiecki2020}, shown as eq.~\ref{eq:HIMS} and eq.~\ref{eq:H2MS}.}
		\label{fig:hims}
	\end{figure*}
	
	In Figure~\ref{fig:hi-mstar}, we show comparisons of \hi-stellar mass relations between observational measurements of G21 (circles and squares) and simulation predictions for SFGs (blue solid lines) and QGs (red dotted lines). From left to right, we present simulation results of Illustris, TNG, and SIMBA, respectively.
	The \hi\ spectra stacking in G21 works by adding the \hi\ fluxes of all stacked galaxies. Even for those galaxies with \hi\ fluxes below the individual detection limit of the telescope, their contribution to the total signal is still counted in the stacking. This is especially important for QGs with typically low \hi\ fluxes \citep{Saintonge2016}. For fair comparisons with G21, we calculate $\langle M_{\rm HI}\rangle$ by $\sum_i M_{{\rm HI},i}/N$, where $M_{{\rm HI},i}$ is the \hi\ mass for the $i$-th galaxy in a given stellar mass bin,  including those with $M_{\rm HI}=0$ (i.e. their \hi\ mass is below the simulation resolution). $N$ is the total number of galaxies in this bin.
	
	It is consistently shown in all simulations that QGs have much less \hi\ gas content than SFGs, i.e. the quenched state of galaxies is associated with the loss of \hi\ reservoir, confirming the finding of G21. 
	\begin{figure*}
		\centering
		\includegraphics[width=0.8\textwidth]{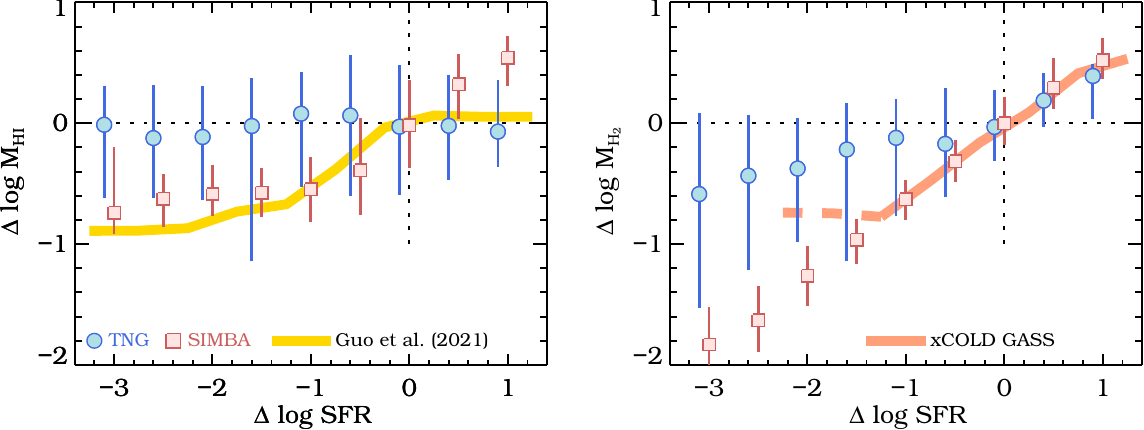}
		\caption{Left: relation between SFR and $M_{\rm HI}$, scaled by the corresponding values on the SFMS, i.e. ${\rm SFR}/{\rm SFR_{MS}}$ and $M_{\rm HI}/M_{\rm HI,MS}$. Right: similar relation between SFR and $M_{\rm H_2}$. The measurements of TNG, SIMBA, G21, and xCOLD GASS are shown as blue circles, red squares, yellow lines, and peach lines, respectively. The peach dashed line in the right panel shows the upper limits of the xCOLD GASS non-detections. The data points shown are the median values and the errors are estimated from the $16^{\rm th}$--$84^{\rm th}$ percentile ranges. For clarity, the SFR measurements for TNG are shifted leftwards by 0.1~dex. The horizontal dotted lines indicate the positions of HIMS (left) and H$_2$MS (right), while the vertical ones are for the SFMS. Only galaxies with $M_\ast>10^{9.5}\msun$ are included in both observations and simulations. }
		\label{fig:deltaHI}
	\end{figure*}
	
	For Illustris simulation, there is almost no dependence of $M_{\rm HI}$ on $M_\ast$ for SFGs, with significantly higher \hi\ mass for $M_\ast<10^{10.5}\msun$ compared to G21. However, the \hi\ mass of QGs are on average 0.4~dex lower than observed values, which is consistent with the findings of \cite{Genel2014}. 

	The physical origin of discrepancies in Illustris is that quasar-mode feedback in low-mass galaxies is less effective while radio-mode is too strong for massive ones. In TNG, the coupling efficiency in the high-accretion state $\epsilon_{\rm f,high}$ has been increased to 0.1 \citep{Weinberger2017}, compared with 0.05 in Illustris. This reduces the \hi\ mass for SFGs with low stellar masses, i.e. $M_\ast<10^{10}\msun$, producing better agreement with observation. But the \hi\ mass of QGs in this range is further decreased, leading to larger discrepancies. For massive galaxies of $M_\ast>10^{10}\msun$, energy released from the kinetic feedback, which is in the form of momentum injection, significantly increases in TNG \citep{Weinberger2018}. This effectively reduces the gas density surrounding the central black hole, but the remaining total \hi\ gas still seems to be overabundant compared to G21. It is surprising that the differences between SFGs and QGs are decreasing for massive galaxies, with QGs possessing far too much \hi, but not triggering star formation.   
	
	SIMBA generally agrees with observations well for both SFGs and QGs, with the difference in $M_{\rm HI}$ remaining roughly constant at around 0.6~dex as in G21. There is a trend in SIMBA that the dependence of $M_{\rm HI}$ on $M_\ast$ becomes much weaker for $M_\ast>10^{10}\msun$, which is especially clear for QGs with the abrupt change of slope occurring at around $M_\ast\sim10^{10.1}\msun$. It is mainly caused by the mode changes in the AGN feedback models of SIMBA, as this mass scale corresponds to a black hole mass around $10^{7.5}\msun$ where the jet mode feedback is starting to take effect and the black hole growth is becoming slower \citep{Habouzit2021}.  
	
	Due to the large discrepancy between Illustris and observation, we will only analyze the TNG and SIMBA simulations in the following and focus on galaxies with $M_\ast>10^{9.5}\msun$.

	\subsection{Star Formation and the Cold Gas Reservoir}
	
	Following G21, we similarly define the \hi\ main sequence (HIMS) as the median values of $M_{\rm HI}(M_\ast)$ in each $M_\ast$ bin for SFGs, shown as the black points in the top panels of Figure~\ref{fig:hims}. The distribution of QGs is represented by the gray dots in each panel. QGs in TNG have comparable amounts of \hi\ gas to SFGs, as seen in Figure~\ref{fig:hi-mstar}. But there is an apparent offset between the \hi\ masses of SFGs and QGs in SIMBA, which is also consistent with right panel in Figure~\ref{fig:hi-mstar}. We also define the H$_2$ main sequence (H$_2$MS) in the same way as the HIMS in the bottom panels of Figure~\ref{fig:hims}, where the distributions of SFGs and QGs are similar to the top panels. For comparison, we also show the observational HIMS from G21, as well as the H$_2$MS from \cite{Janowiecki2020} as red dotted lines in Figure~\ref{fig:hims}, which are defined as, 
	\begin{eqnarray}
		\log({\rm M_{HIMS}}/\msun)&=&0.42\log{\rm M_\ast}+5.35 \label{eq:HIMS} \\
		\log({\rm M_{H_2MS}}/\msun)&=&0.74\log{\rm M_\ast}+1.44 \label{eq:H2MS}.	
	\end{eqnarray}
   Because the HIMS and H$_2$MS in simulations cannot be simply described by power-law relations, we will use interpolation in the following calculations. The HIMS relations in both TNG and SIMBA agree with that of G21 reasonably well. But $M_{\rm HI}$ is over-predicted in TNG for massive galaxies with $M_\ast>10^{10.5}\msun$, as in Figure~\ref{fig:hi-mstar}. The discrepancies are relatively smaller here as the black dots shown are the median values rather than $\langle M_{\rm HI}\rangle$ in Figure~\ref{fig:hi-mstar}. 
	
	However, both TNG and SIMBA over-predict the H$_2$MS by around 0.5~dex, as also seen in \cite{Dave2020}. In this sense, it is more meaningful and practical to compare the distances from corresponding main sequence values for SFR, $M_{\rm HI}$ and $M_{\rm H_2}$, as follows,
	\begin{eqnarray}
		\Delta\log{\rm SFR}&=&\log{\rm SFR}-\log{\rm SFR_{MS}} \\
		\Delta\log{\rm M_{HI}}&=&\log{\rm M_{HI}}-\log{\rm M_{HIMS}} \\
		\Delta\log{\rm M_{H_2}}&=&\log{\rm M_{H_2}}-\log{\rm M_{H_2MS}} .	
	\end{eqnarray}
	
    The advantage of using these scaled measurements is the removal of stellar mass dependence. The traditional measurements of sSFR and the gas fraction are decreasing with stellar mass, even when galaxies are still on the SFMS \citep{Saintonge2016, Catinella2018}. Therefore, the trends of decreasing sSFR and gas fractions with other physical parameters would potentially be complicated by the stellar mass dependence.
	
	In the left panel of Figure~\ref{fig:deltaHI}, we quantitatively compare the relations between $\Delta\log{\rm SFR}$ and $\Delta\log{\rm M_{HI}}$. The predictions from TNG and SIMBA are compared with the observational measurements of G21. For fair comparisons, we only include galaxies with $M_\ast>10^{9.5}\msun$ in both observations and simulations. As expected, when the SFR decreases, galaxies in SIMBA behave similarly to the observational measurements, with $M_{\rm HI}$ smoothly reduced by $\sim0.6$~dex during quenching. But it seems to over-predict $M_{\rm HI}$ for galaxies above the SFMS, which are dominated by low-mass galaxies of $M_\ast<10^{10.5}\msun$. However, the \hi\ mass distribution in TNG has a much larger scatter and the median $M_{\rm HI}$ only marginally decreases by less than 0.2~dex, even though the SFR decreases by more than 2~dex from the SFMS. 
	
	The trend is very similar if we instead compare the relation between SFR and H$_2$ for the two simulations, shown in the right panel of Figure~\ref{fig:deltaHI}. Obtaining H$_2$ measurements for a large number of galaxies (as for the HI measurements from G21) is very challenging in technique, especially for QGs. Because of the low line transition probability of H$_2$, the CO emission lines are commonly considered as the tracers of H$_2$. But the lower abundance of CO in galaxies makes it also harder to detect, compared to the abundant \hi. We show the H$_2$ measurements from the xCOLD GASS sample \citep{Saintonge2017} of 532 galaxies (199 are non-detections) for comparison, and adopt the SFMS and H$_2$MS definitions for xCOLD GASS galaxies from \cite{Janowiecki2020}. We note that the H$_2$ measurements for $\Delta\log{\rm SFR}<-1.2$ are mostly upper limits due to the non-detections (shown as the peach dashed line in the right panel). Therefore, QGs have $M_{\rm H_2}$ decreased by at least 0.7~dex from the main sequence values, which is consistent with the \hi\ observations of G21. 
	
	The agreement between SIMBA and xCOLD GASS is remarkably good for the available measurements of $\Delta\log{\rm SFR}>-1.2$. TNG shows a similar agreement for SFGs of $\Delta\log{\rm SFR}>-0.5$. For lower SFRs, the median values in TNG are slightly higher than the xCOLD GASS measurements, but they are still consistent within the estimated errors. We will further discuss the effect of different aperture sizes when comparing the measurements from simulations with those from observations in Section~\ref{subsec:sigma1}. 
	\begin{figure*}
		\centering
		\includegraphics[width=0.8\textwidth]{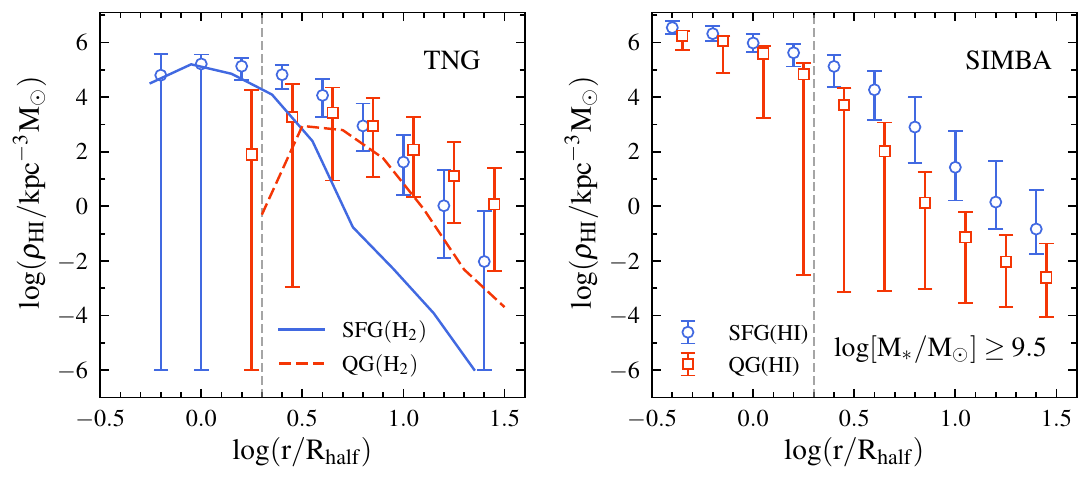}
		\caption{3D \hi\ mass density profile $\rho_{\rm HI}(r/R_{\rm half})$ for TNG (left) and SIMBA (right), where the distance to galaxy center $r$ is scaled by the corresponding stellar half mass radius $R_{\rm half}$. SFGs and QGs with $M_\ast\ge10^{9.5}\msun$ are represented by blue circles and red squares, respectively. We show the median values along with the $20^{\rm th}$--$80^{\rm th}$ percentile ranges. 3D H$_2$ mass density profiles in TNG are also included in the left panel, shown as blue solid (SFGs) and red dotted (QGs) lines. The vertical dashed line in each panel indicates the position of $r=2R_{\rm half}$, which is used to compute the gas masses within the galaxy radii.}
		\label{fig:profile}
	\end{figure*}
	\begin{figure*}
		\centering
		\includegraphics[width=0.8\textwidth]{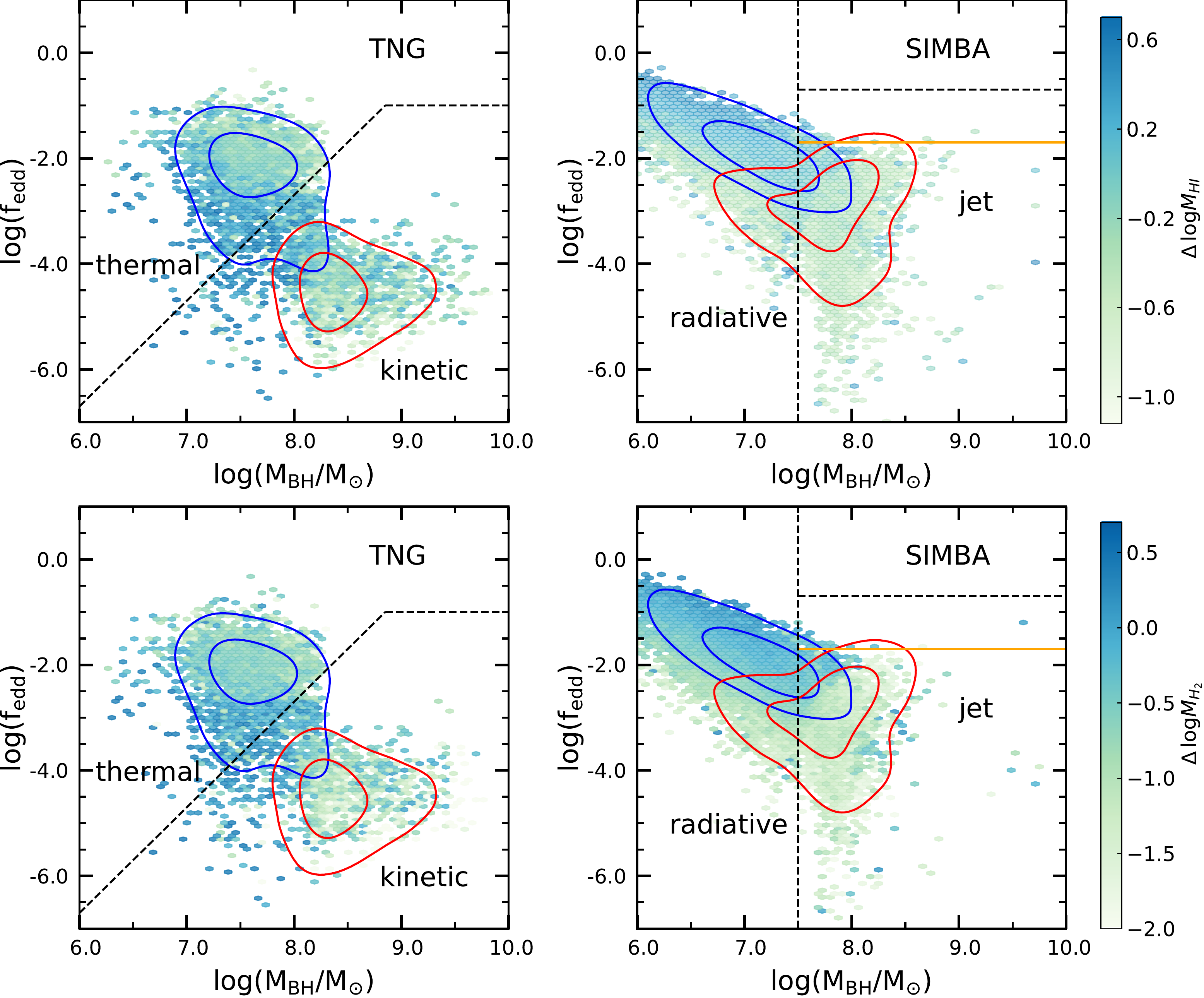}
		\caption{Distribution of galaxies as a function of $M_{\rm BH}$ and $f_{\rm edd}$ for TNG (left) and SIMBA (right). The blue and red contours show the distributions of SFGs and QGs, respectively. The contour levels are the $50^{\rm th}$ and $90^{\rm th}$ percentiles. The galaxy distributions are color coded by $\Delta\log M_{\rm HI}$ (top panels) and $\Delta\log M_{\rm H_2}$ (bottom panels), respectively. The demarcation lines for different AGN feedback modes are displayed as dotted lines. The orange solid line in the right panel shows the position of $f_{\rm edd}=0.02$ where the maximal ejection velocity is achieved.}
		\label{fig:mbh_fedd}
	\end{figure*}
	
	\subsection{Cold Gas Density Profile}\label{sec:profile}
	
	In addition to the total cold gas content for SFGs and QGs, it is useful to compare the spatial distribution of \hi\ in the two populations. In Figure~\ref{fig:profile}, we show the 3D density profiles of \hi\ for galaxies with $M_\ast\ge10^{9.5}\msun$ in TNG (left) and SIMBA (right). The median 3D \hi\ mass densities $\rho_{\rm HI}$ for SFGs and QGs are represented by blue circles and red squares, respectively. We show the median values along with the $20^{\rm th}$--$80^{\rm th}$ percentile ranges. There are 4716 (10729) SFGs and 1638 (5593) QGs with $M_\ast\ge10^{9.5}\msun$ in TNG (SIMBA). $\rho_{\rm HI}$ is obtained from the median \hi\ densities of all SFGs (or QGs) in equally spaced $\log(r/R_{\rm half})$ bins, where the distance to galaxy center $r$ is scaled by the corresponding stellar half mass radius $R_{\rm half}$.
	
	We note that there are still many galaxies with $\rho_{\rm HI}$ below the simulation resolutions (i.e. $M_{\rm HI}=0$) in each $r/R_{\rm half}$ bin (especially for QGs), and we have manually set the lower limit as $\rho_{\rm HI}=10^{-6}\,\msun/{\rm kpc}^{3}$ to properly show these galaxies in the figure. The percentile range is selected to avoid including too many galaxies at the lower limit. For the QGs in TNG, the fraction of galaxies with $M_{\rm HI}=0$ in the two innermost bins are 75.82\% and 56.96\%, respectively. Thus, we do not show the median values for these two bins in the left panel of Figure~\ref{fig:profile}.
	
	There are stark differences in the \hi\ density distributions of SFGs and QGs in TNG and SIMBA. QGs in TNG have significantly lower $\rho_{\rm HI}$ within $\sim3R_{\rm half}$, but there is much more \hi\ gas in the outer regions compared to their star-forming counterparts. However, QGs in SIMBA have consistently lower \hi\ mass than the corresponding SFGs. In the inner region of $r<3R_{\rm half}$, QGs seem to have close \hi\ distributions to SFGs. This is very different from the trend in TNG, where QGs have significantly lower \hi\ densities in the inner, but much higher in the outer. 
	
	We also check the profile of H$_2$ density in TNG and find a very similar trend as in the case of \hi, shown as blue solid and red dashed lines for SFGs and QGs, respectively. But most of the median values of $\rho_{\rm H_2}$ in each $r/R_{\rm half}$ bin in SIMBA is below the simulation resolution, which might be caused by the discrete distributions of H$_2$ clumps in SIMBA as they are calculated on-the-fly at each time-step \citep{Dave2020}, rather than the post-processing treatment in TNG. Therefore, we do not show the $\rho_{\rm H_2}$ for SIMBA. But if we calculate the average $\rho_{\rm H_2}(r/R_{\rm half})$ in each bin, the trends in both simulations will be quite similar as in the case of $\rho_{\rm HI}$.
		
	\begin{figure*}
		\centering
		\includegraphics[width=0.75\textwidth]{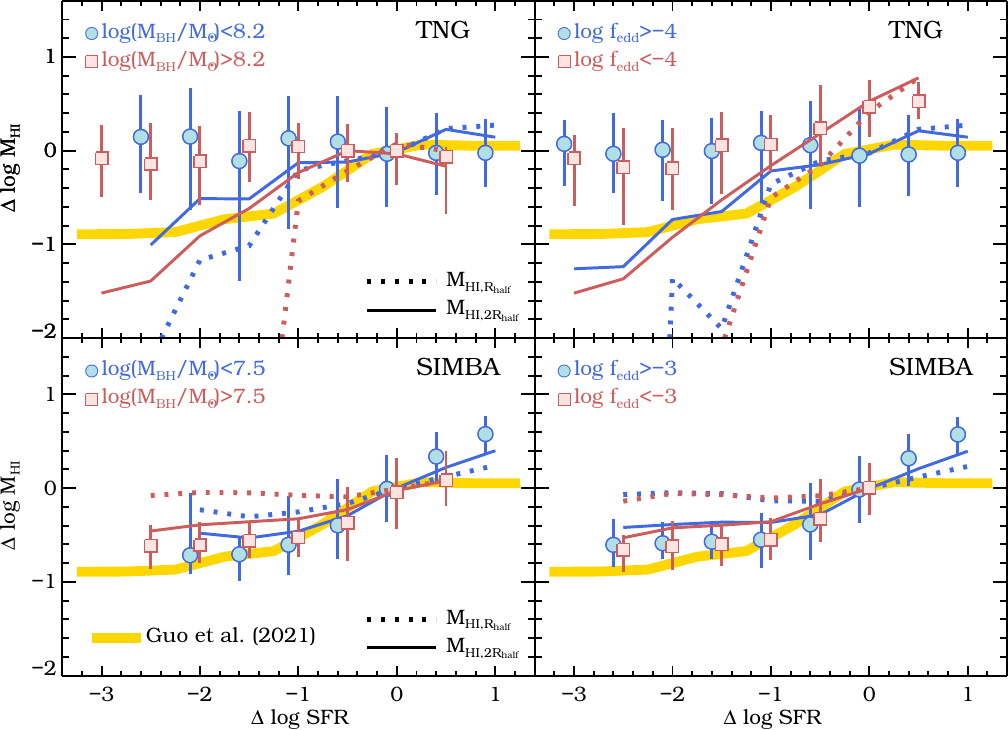}
		\caption{Relation between $\Delta\log{\rm SFR}$ and $\Delta\log{\rm M_{ HI}}$ for TNG (upper panels) and SIMBA (bottom panels). The galaxy samples are separated into two $M_{\rm BH}$ (left) and $f_{\rm edd}$ (right) bins, shown as different colors. The filled circles and squares are for the total \hi\ mass $M_{\rm HI}$ as in Figure~\ref{fig:deltaHI}, while solid and dotted lines represent the corresponding \hi\ mass measurements within $2R_{\rm half}$ and $R_{\rm half}$, denoted as $M_{\rm HI,2R_{\rm half}}$ and $M_{\rm HI,R_{\rm half}}$ respectively.}
		\label{fig:hims_bh}
	\end{figure*}
	\begin{figure*}
		\centering
		\includegraphics[width=0.75\textwidth]{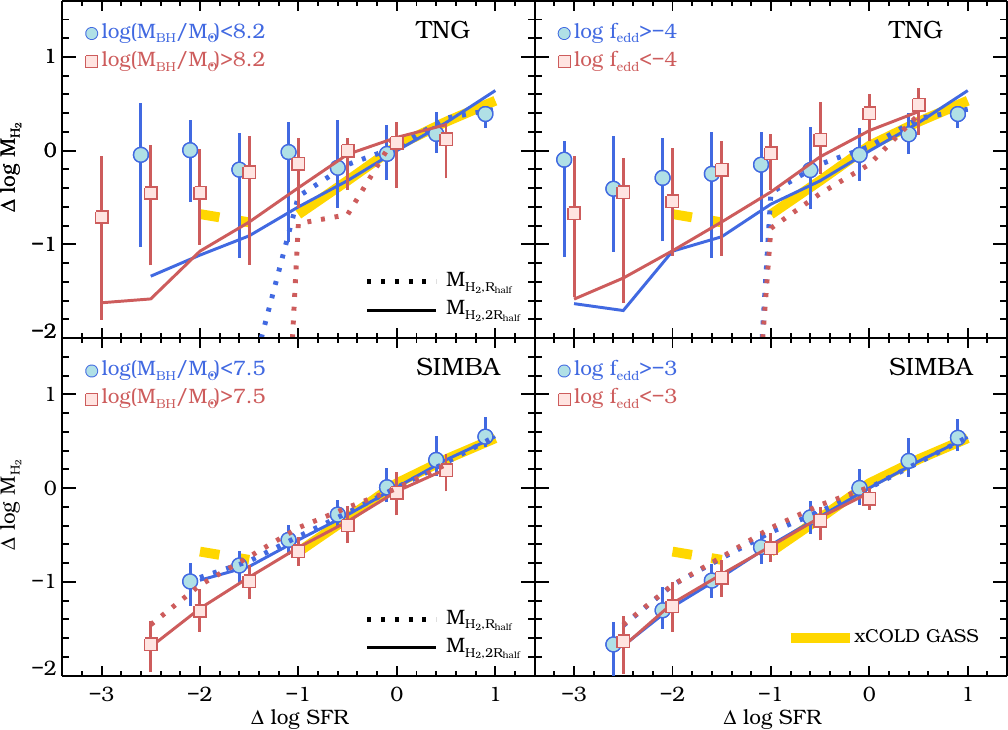}
		\caption{Similar to Figure~\ref{fig:hims_bh}, but for the H$_2$ gas. We also show the observational measurements of xCOLD GASS for comparison.}
		\label{fig:h2ms_bh}
	\end{figure*}
	
	\subsection{Black Hole Growth and AGN feedback}\label{sec:bh}
	While in both TNG and SIMBA, AGN feedback is set as the primary mechanism to quench massive galaxies, it has been suggested that the cumulative energy release from AGN, rather than the instantaneous feedback, determines whether or not a galaxies would be quenched \citep{Terrazas2020,Zinger2020,Piotrowska2022}. It is thus important to check the effects of both cumulative and instantaneous AGN feedback on the cold gas content. 
	Figure~\ref{fig:mbh_fedd} shows the distributions of galaxies in the $M_{\rm BH}$--$f_{\rm edd}$ plane for TNG (left) and SIMBA (right). The galaxy distributions are color coded by $\Delta\log M_{\rm HI}$ (top panels) and $\Delta\log M_{H_2}$ (bottom panels). The distributions of central SFGs and QGs are represented by blue and red contours, respectively. The demarcation lines for different AGN feedback modes (detailed in Sections 2.3 and 2.4) in both simulations are shown as dotted lines.  
	
	QGs in TNG can be well-separated from SFGs with a black hole mass threshold of $M_{\rm BH}>10^{8.2}\msun$. However, SFGs and QGs in SIMBA are more overlapped in the $M_{\rm BH}$--$f_{\rm edd}$ plane. But the cut of $M_{\rm BH}=10^{7.5}\msun$ can still reasonably separate the two populations, as well as the two AGN feedback modes. They can also be distinguished with the Eddington ratio cut of $\log f_{\rm edd}\sim-3$ \citep{Thomas2019}. The orange solid line in the right panel shows the position of $f_{\rm edd}=0.02$, where the maximal ejection velocity is achieved in SIMBA. It is clear that QGs in SIMBA mostly have $f_{\rm edd}<0.02$. The values of $\Delta\log M_{\rm HI}$ and $\Delta\log M_{H_2}$ are also gradually decreasing from SFGs to QGs, consistent with the results of Figure~\ref{fig:deltaHI}.
	
	Unlike SIMBA, gas-rich galaxies in TNG tend to distribute more in the transition area between the thermal and kinetic modes. It is likely caused by the fact that both strong thermal and kinetic feedback will significantly reduce the cold gas reservoir. But the distribution of gas-rich galaxies is still wide-spread in TNG, resulting in the weak dependence on SFR in Figure~\ref{fig:deltaHI}.
	
	In Figure~\ref{fig:hims_bh}, we show a similar relation between $\Delta\log{\rm SFR}$ and $\Delta\log{\rm M_{ HI}}$ as in Figure~\ref{fig:deltaHI} for TNG (upper panels) and SIMBA (bottom panels). The galaxy samples are further separated into two $M_{\rm BH}$ (left) and $f_{\rm edd}$ (right) bins, shown as different colors. The filled circles and squares are for the total \hi\ mass $M_{\rm HI}$ as in Figure~\ref{fig:deltaHI}, while the solid and dotted lines represent the corresponding \hi\ mass measurements within $2R_{\rm half}$ and $R_{\rm half}$, denoted as $M_{\rm HI,2R_{\rm half}}$ and $M_{\rm HI,R_{\rm half}}$, respectively. 
	
	The measurements of $M_{\rm HI,R_{\rm half}}$, $M_{\rm HI,2R_{\rm half}}$ and $M_{\rm HI}$ represent the \hi\ gas within the galactic center, the whole galaxy and the whole subhalo (or circumgalactic medium), respectively. The three $\Delta\log{\rm M}$ are measured as the distances to their own HIMS. It demonstrates that the depletion of \hi\ gas with larger $M_{\rm BH}$ in TNG is increasingly stronger when approaching the galaxy center. The overall trend of $M_{\rm HI,2R_{\rm half}}$ agrees much better with the observation. Because $\Delta\log{\rm M_{ HI}}$ does not reduce strongly with the decreasing SFR, it means that the kinetic AGN feedback in TNG mainly works as redistributing the inner gas to the outer that prevents the star formation in the galactic center.
		
	The effect of $f_{\rm edd}$ is very weak for $\Delta\log{\rm SFR}<-1$. But for SFGs (i.e. $\Delta\log{\rm SFR}>-1$), galaxies with high $f_{\rm edd}$ will have somewhat less \hi\ gas, by a maximal amount of $\sim0.5$~dex, which is related to the thermal energy release in the high-accretion model AGN feedback \citep{Weinberger2017}. It is also clear that the influence of $f_{\rm edd}$ does not change with the distance to galactic center.
	 
	The measurements of $\Delta\log{\rm M_{\rm HI}}$ within different radii are all consistent with each other for SFGs in cases of different $M_{\rm BH}$ and $f_{\rm edd}$. We further check that the density profiles of SFGs with different $M_{\rm BH}$ and $f_{\rm edd}$ have similar shapes. It means that the feedback from high-$f_{\rm edd}$ AGNs affects the inner and outer regions equivalently. The thermal energy released in the inner region might be quickly balanced by the \hi\ inflow from outer, causing the overall reduction of $\rho_{\rm HI}$. 
	
	In SIMBA, the trend of $\Delta\log{\rm M_{\rm HI}}$ with the distance to galactic center is contrary to that of TNG. The \hi\ gas in inner region is less changed with SFR, which is also seen in the \hi\ density profiles of Figure~\ref{fig:profile}.  For a given distance to galactic center, there is a trend that galaxies with more massive $M_{\rm BH}$ show slightly weaker dependence of $\Delta\log{\rm M_{\rm HI}}$ on SFR. It is likely related to the smaller mass loading factor of outflows in the jet-mode kinetic feedback \citep{Dave2019}. Similar to TNG, the effect of $f_{\rm edd}$ is also very weak for QGs in SIMBA. Even for the SFGs, there is no strong differences between high and low $f_{\rm edd}$, because the high-accretion mode feedback is also expressed as kinetic energy release. 
	
	The behavior of H$_2$ gas is very similar to \hi\ in TNG, as shown in the top panels of Figure~\ref{fig:h2ms_bh}. The dependence of $M_{\rm H_2}$ on $M_{\rm BH}$ is slightly stronger. It is related to the fact that H$_2$ gas distribution is more concentrated than \hi\ (as it is formed at higher-density regions) and thus more affected by AGN feedback. However, in SIMBA, there is no any strong dependence of the $\Delta\log{\rm SFR}$--$\Delta\log{\rm M_{H_2}}$ relation on $M_{\rm BH}$, $f_{\rm edd}$, or the aperture size of calculating $M_{\rm H_2}$, since the SFR in SIMBA is directly measured from H$_2$ gas as SFR$\propto\rho_{\rm H_2}^{1.5}$. In this sense, the \hi\ gas is less correlated with SFR, as they should be converted to H$_2$ before forming stars, mimicking the observations \citep{Bigiel2008,Leroy2008}. While the \hi/H$_2$ modeling methods in TNG is different. The star formation in TNG is based on the total neutral hydrogen gas density ($n_{\rm H}$) and $f_{\rm H_2}$ is derived from post-processing.

\begin{figure*}
		\centering
		\includegraphics[width=0.8\textwidth]{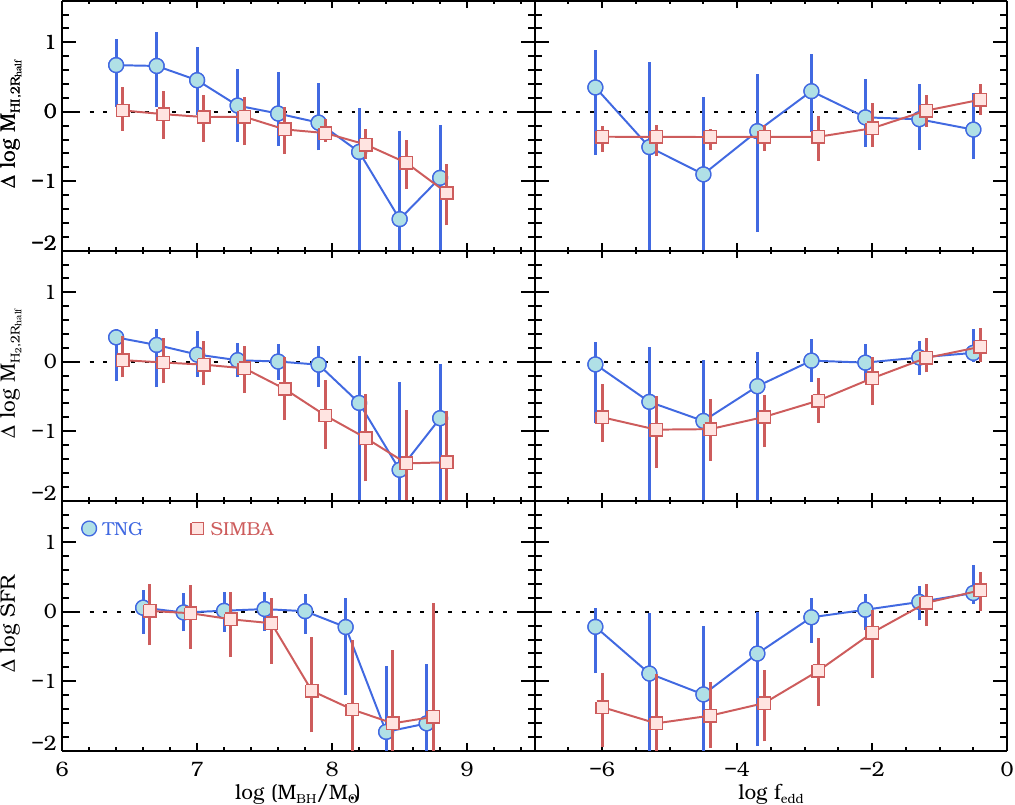}
		\caption{Dependences of $\Delta\log{\rm M_{HI,2R_{\rm half}}}$ (top panels), $\Delta\log{\rm M_{H_2,2R_{\rm half}}}$ (middle panels) and $\Delta\log{\rm SFR}$ (bottom panels) on $M_{\rm BH}$ (left columns) and $f_{\rm edd}$ (right columns), for TNG (blue circles) and SIMBA (red squares).}
		\label{fig:himsdisk}
	\end{figure*}

	As the gas masses within $2R_{\rm half}$ are more reasonable quantities to reflect the effect of AGN feedback, we will focus on the corresponding values in the following. In Figure~\ref{fig:himsdisk}, we show the dependence of $\Delta\log{\rm M_{HI,2R_{\rm half}}}$ (top panels), $\Delta\log{\rm M_{H_2,2R_{\rm half}}}$ (middle panels) and $\Delta\log{\rm SFR}$ (bottom panels) on $M_{\rm BH}$ (left columns) and $f_{\rm edd}$ (right columns), for both TNG and SIMBA. The SFR starts to drop below the main sequence when $M_{\rm BH}$ increases above the input threshold values for jet or kinetic mode AGN feedback. The transition from radiative to jet modes is accompanied with the smaller $f_{\rm edd}$, along with the decreasing of $M_{\rm HI}$ and $M_{\rm H_2}$. For QGs with $\Delta\log{\rm SFR}<-1$, there is only weak dependence of $\Delta\log{\rm M_{HI,2R_{\rm half}}}$ and $\Delta\log{\rm M_{H_2,2R_{\rm half}}}$ on $f_{\rm edd}$, while they both continue to decrease with increasing $M_{\rm BH}$. We also note that both $M_{\rm HI,2R_{\rm half}}$ and $M_{\rm H_2,2R_{\rm half}}$ in TNG decrease with $M_{\rm BH}$ even before the kinetic-mode feedback kicks in, when galaxies are still on the SFMS. It supports the idea that cumulative, rather than instantaneous, AGN feedback drives the quenching. 
	
	The complex correlation among $M_{\rm BH}$, $f_{\rm edd}$ and SFR makes it not straightforward to use correlation coefficients for investigating the dominant driver of gas depletion and quenching. \cite{Piotrowska2022} applied the random forest analysis in SDSS galaxies and several hydrodynamical simulations (including Illustris and TNG). They concluded that black hole mass is the most predictive parameter for the change of SFR. The dependence of SFR on $\dot M_{\rm BH}$ becomes much weaker when accounting for its correlation with $M_{\rm BH}$. It complements our analysis here that the cold gas depletion is also driven by the black hole growth. The three parameters of $M_{\rm HI,2R_{\rm half}}$ (or $M_{\rm H_2,2R_{\rm half}}$), SFR, and $M_{\rm BH}$ would evolve altogether to form the results shown in Figures~\ref{fig:hims_bh} and~\ref{fig:h2ms_bh} \citep{Cui2021}.
	\begin{figure*}
		\centering
		\includegraphics[width=0.8\textwidth]{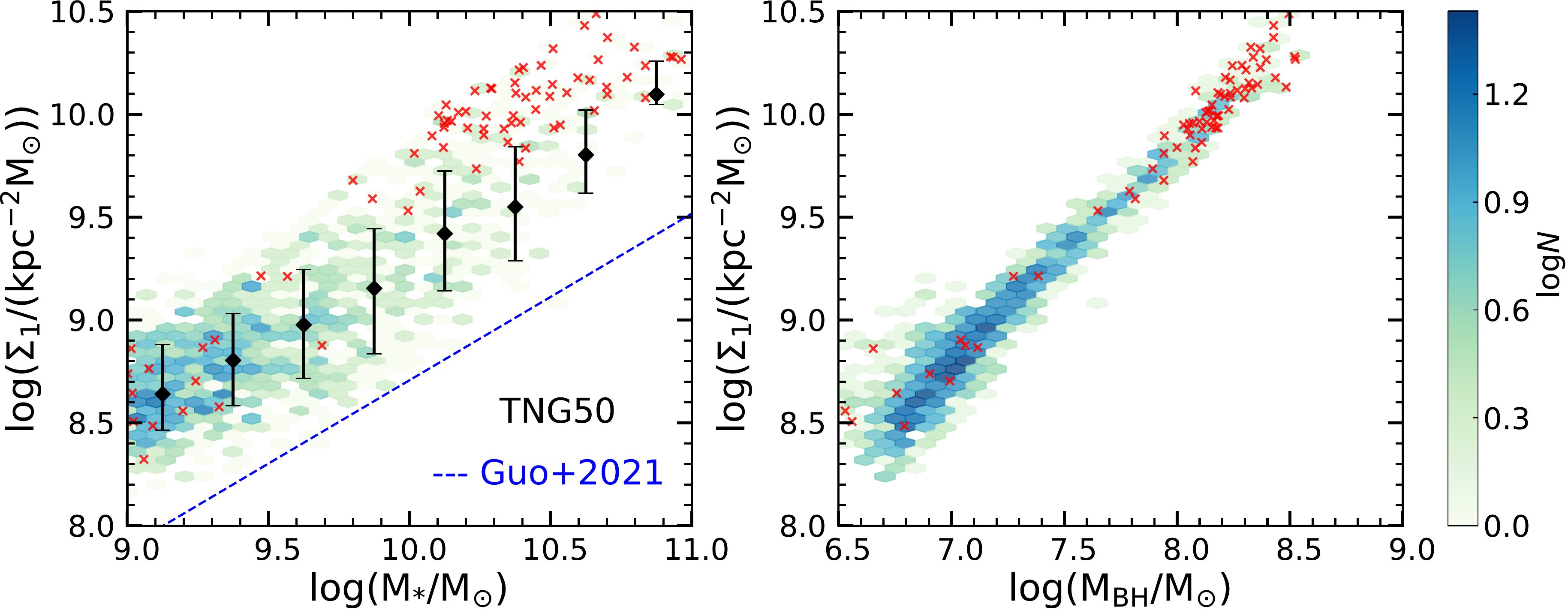}
		\caption{Distribution of $\Sigma_1$ as functions of $M_\ast$ (left panel) and $M_{\rm BH}$ (right panel) in TNG50. QGs are showed as red crosses in each panel. The black symbols with errors in the left indicate the $\Sigma_{\rm 1,MS}$ for SFGs in TNG50. The blue dashed line represents the best-fitting observational measurements of $\Sigma_{\rm 1,MS}$ in G21 ($\log\Sigma_{\rm 1,MS}=0.81\log M_\ast+0.607$). The probability distribution of all galaxies is represented by the logarithmic color scales.}
		\label{fig:sigma1ms}
	\end{figure*}
	
	\subsection{Connection to the Central Stellar Surface Density}\label{subsec:sigma1}
	It has been found in G21 that the \hi\ reservoir will decrease with the central stellar surface density within 1kpc (denoted as $\Sigma_1$), as $M_{\rm HI}\propto\Sigma_1^{-2}$ when galaxies start to quench. It is explained as a signature of quenching driven by the compaction events \citep{Dekel2014,Zolotov2015} in G21. In this scenario, violent disc instability caused by the intense cold gas inflow would result in the creation of dense cores, which then give rises to a phase of high SFRs \citep{Dekel2014}. The rapid consumption of cold gas into stars, along with the gas outflow due to efficient stellar feedback, could cause the dramatic decrease of the cold gas reservoir. In the meantime, the gas inflow also promotes the growth of central black hole, which would lead to the cold gas depletion with energetic AGN feedback. The decrease of the cold gas surface density also help stabilize the disk that potentially prevents the gas from forming stars. When the host halos of galaxies grow beyond a typical mass of $10^{12}\msun$, the virial shock heating will further take effect to shut down the cold gas supply, causing the starvation (see more discussions in G21). \cite{Chen2020} proposed a phenomenological model that the black hole growth in star-forming galaxies scales with $\Sigma_1$, as $M_{\rm BH}\propto\Sigma_1^{1.76}$. It predicts that majority of the black hole mass assembly happen when galaxies are in the compact star-forming phase, which would then drive the quenching. It is not yet clear which mechanism plays the dominant role in the galaxy quenching, but agreement has been reached that compactness is a necessary, though not sufficient, condition for the quenching to happen \citep{Cheung2012,Fang2013,Zolotov2015,Barro2017}.
	
	We employ the TNG50-1 (hereafter TNG50) simulation to investigate the correlation between $\Sigma_1$ and \hi\ gas, since its gravitational softening length is only 288~pc, compared to 738~pc in TNG100-1 \citep{Pillepich2019}. We directly calculate $\Sigma_1$ in TNG50 by rotating the axes of galaxies to the face-on direction according to the moment of stellar inertia tensor within $2R_{\rm half}$ and summing up all stellar particles within the projected distance of 1~kpc. We show the dependence of $\Sigma_1$ on $M_\ast$ (left) and $M_{\rm BH}$ (right) in Figure~\ref{fig:sigma1ms}. The black symbols with errors indicate the median $\Sigma_1$ for SFGs, which is defined as the $\Sigma_1$ main sequence, $\Sigma_{\rm 1,MS}$. The blue dashed line represents the best-fitting observational measurements of $\Sigma_{\rm 1,MS}$ in G21 ($\log\Sigma_{\rm 1,MS}=0.81\log M_\ast+0.607$). The measurements in TNG50 are consistent with the observation that $\Sigma_1$ is generally increasing with $M_\ast$, and those massive QGs (represented by crosses) typically have higher $\Sigma_1$ than their star-forming counterparts. 
	
	Although $\Sigma_{\rm 1,MS}$ in TNG50 is systematically higher than the observation of G21 by about 0.6~dex (as also seen similarly in \citealt{Varma2022}), they have a very similar slope of 0.81. It makes the simulation and observation more comparable when we define the relative difference to $\Sigma_{\rm 1,MS}$ as 
	\begin{equation}
		\Delta\log\Sigma_1=\log\Sigma_1-\log\Sigma_{\rm 1,MS},
	\end{equation}
	similar to the definition of $\Delta\log{\rm SFR}$ above.
	
	Compared to the scattered $\Sigma_1$--$M_\ast$ distribution, the relation between $\Sigma_1$ and $M_{\rm BH}$ is much tighter, with $\Sigma_1\propto M_{\rm BH}$. The average scatter in $\Sigma_1$ is only 0.08~dex for $M_{\rm BH}>10^{7.2}\msun$, which means that $\Sigma_1$ is an excellent indicator of $M_{\rm BH}$ for massive galaxies in TNG50, supporting the model of \cite{Chen2020}. However, the slope predicted by the theoretical model of \cite{Chen2020} is around $\Sigma_1\propto M_{\rm BH}^{0.568}$, and the scatter there is also much larger. It indicates that the black hole mass growth with the stellar density is much slower in TNG50.
	
	In the $\Sigma_1$--$M_\ast$ relation, the scatter is largely caused by the additional dependence of $\Sigma_1$ on SFR, with QGs having higher $\Sigma_1$ \citep[see e.g.,][]{Fang2013,Barro2017}. Therefore, the tight correlation between $\Sigma_1$ and $M_{\rm BH}$ implies that the dependence of $\Sigma_1$ on SFR stems from the black hole growth. 
	
\begin{figure*}
		\centering
		\includegraphics[width=1.0\textwidth]{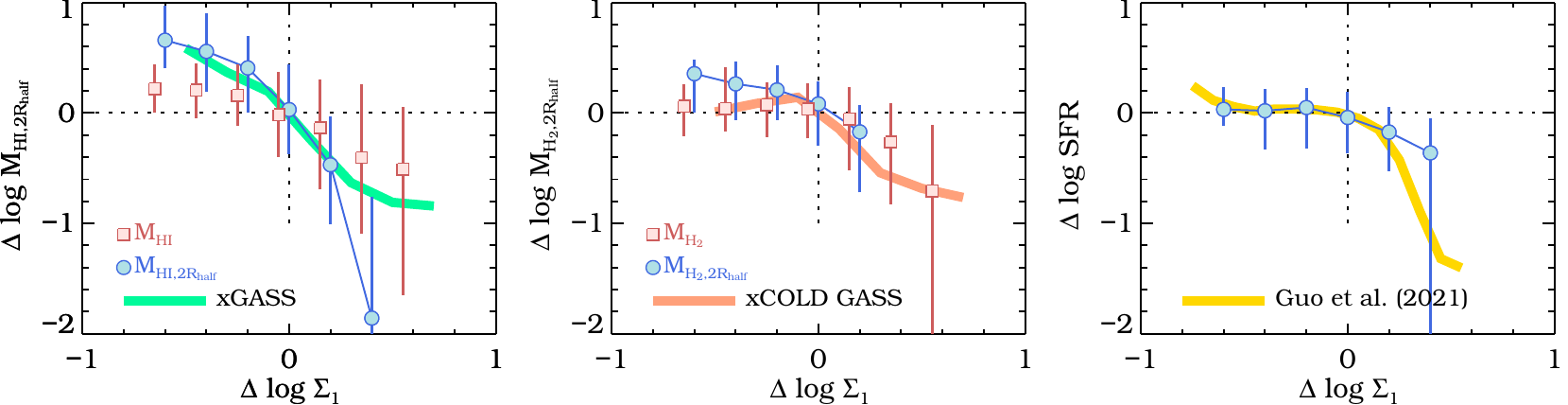}
		\caption{Similar to Figure~\ref{fig:himsdisk}, but for the dependences of $\Delta\log{\rm M_{HI,2R_{\rm half}}}$ , $\Delta\log{\rm M_{H_2,2R_{\rm half}}}$ and $\Delta\log{\rm SFR}$ on $\Delta\log\Sigma_1$ in TNG50. The measurements of TNG50 (circles) are compared with the observations of xGASS, xCOLD GASS and G21. We also show the corresponding measurements of $M_{\rm HI}$ and $M_{\rm H_2}$ in TNG50 as the squares for comparison. }
		\label{fig:hisigma1}
	\end{figure*}

	Similar to Figure~\ref{fig:himsdisk}, we show in Figure~\ref{fig:hisigma1} the dependences of $\Delta\log{\rm M_{HI,2R_{\rm half}}}$, $\Delta\log{\rm M_{H_2,2R_{\rm half}}}$ and $\Delta\log{\rm SFR}$ on $\Delta\log\Sigma_1$. The measurements in TNG50 (filled circles) are compared with the observations of xGASS \citep{Catinella2018} for $M_{\rm HI}$, xCOLD GASS \citep{Saintonge2017} for $M_{\rm H_2}$ and G21 for SFR. We obtain the $\Sigma_1$ measurements for galaxies in xGASS and xCOLD GASS by matching with the galaxy sample of G21. We use the \hi\ measurements of xGASS instead of G21, as the \hi\ stacking in G21 is not made explicitly for the $\Delta\log\Sigma_1$ bins. We also note that the observational measurements of xGASS and xCOLD GASS are for the total $M_{\rm HI}$ and $M_{\rm H_2}$. 
	
	The \hi\ line fluxes in xGASS were observed using the Arecibo telescope with a beam size of 3.5\arcmin \citep{Catinella2010,Catinella2018}, while the H$_2$ measurements in xCOLD GASS were inferred from the CO(1-0) line luminosity using the IRAM 30m telescope with a beam size of 22\arcsec \citep{Saintonge2017}. The H$_2$ gas is primarily located in the ISM, but the distribution of \hi\ gas can be more extended in the galactic halo. Integrated measurements of $M_{\rm HI}$ and $M_{\rm H_2}$ are thus evaluated at different aperture sizes, which might not correspond to relevant scales in the simulations. Given that both the total \hi\ and H$_2$ mass functions, as well as their density profiles, in simulations are in reasonable agreement with observations \citep{Diemer2019,Dave2020}, fair comparisons should be made between observations and the overall measurements of \hi\ and H$_2$. However, as shown in the previous figures, the trends using the measurements within $2R_{\rm half}$ agree much better with observations, which would then better reflect the inherent physics. Similar comparisons based on the \hi\ estimates within optical disks have been presented in \cite{Wang2020}.
	
	These measurements in TNG50 are in good agreement with observations, except that the decreasing trend of SFR with $\Sigma_1$ is much shallower in TNG50, which is likely caused by resolution effects and the less sampling of massive galaxies \citep{Donnari2021}. These dependences on $\Delta\log\Sigma_1$ are very similar to those on $M_{\rm BH}$ presented in Fig.~\ref{fig:himsdisk}, which further suggests that the observed decreasing of $M_{\rm HI}$, $M_{\rm H_2}$ and SFR with increasing $\Sigma_1$ is a reflection of the black hole growth. However, if we instead use $\Delta\log{\rm M_{HI}}$ and $\Delta\log{\rm M_{H_2}}$ in TNG50 (shown as the squares), the dependences on $\Sigma_1$ would become much weaker, similar to the results shown in Figs.~\ref{fig:hims_bh} and~\ref{fig:h2ms_bh}.
	
	We also compare galaxy distributions in the $\Sigma_1$--$f_{\rm edd}$ plane and find very weak correlation between the two, which is as a result of the tight correlation between $\Sigma_1$ and $M_{\rm BH}$, as well as the weak correlation between $M_{\rm BH}$ and $f_{\rm edd}$, as expected from Figure~\ref{fig:mbh_fedd}. Combining all the results above, it suggests that the cumulative energy release from the black hole growth is the dominant reason driving the cold gas depletion and galaxy quenching. The observations of cold gas content can then be used to constrain the AGN feedback mechanisms.

    \section{Discussion}\label{sec:discussion}

    \subsection{Post-processing versus On-the-fly Methods}\label{subsec:hi method}
    As the decomposition of total neutral hydrogen into \hi\ and H$_2$ is very different in TNG and SIMBA, caution should be taken when interpreting the cause of quenching. More specifically, $M_{\rm H_2}$ is post-processed in TNG from multiplying the neutral hydrogen mass in each gas cell by the molecular fraction $f_{\rm H_2}$ in Equation~(\ref{eq:fh2}) using the input SFR, which is also determined from the neutral hydrogen density ($n_{\rm H}$). In this sense, neither \hi\ nor H$_2$ is directly responsible for the change of SFR. The density distributions of \hi\ and H$_2$ in TNG are quite similar, as shown in Figure~\ref{fig:profile}. All other measurements of \hi\ and H$_2$ for TNG galaxies in Figures~\ref{fig:hims_bh}--\ref{fig:himsdisk} have similar trends as well.  

    However, the \hi\ and H$_2$ masses are independently modeled in SIMBA and calculated on the fly with the simulation run. The decrease of SFR is simply caused by the reduction of molecular gas content (Equation~\ref{eq:sfr_simba}), as manifested in the bottom panels of Figure~\ref{fig:h2ms_bh} for the H$_2$ masses measured in different aperture sizes. It is then possible that the star formation could be quenched while there is still relatively abundant \hi\ gas, by reducing the conversion efficiency from \hi\ to H$_2$. It is the case in inner regions (within $R_{\rm half}$) of QGs in SIMBA, where QGs with very low SFRs can have similar $M_{\rm HI,R_{half}}$ as the SFGs, as seen in Figures~\ref{fig:profile}, ~\ref{fig:hims_bh} and~\ref{fig:h2ms_bh}.

    To conclude, for a given gas cell in TNG, its low SFR is caused by the low $n_{\rm H}$, which indicates both low \hi\ and H$_2$ masses. However, in SIMBA, it only infers a low H$_2$ mass of the gas cell, which not necessarily has a low \hi\ mass. Such differences in the modeling methods of \hi\ and H$_2$ gas can sometimes cause misinterpretation of the results and should be treated with care.

    \subsection{AGN Feedback Mechanisms}
    Despite the different models of black hole growth and feedback in TNG and SIMBA, the shapes of their \hi\ gas density profiles for SFGs are quite similar, as seen in Figure~\ref{fig:profile}. The apparent differences in the density profiles of QGs are mainly caused by the kinetic feedback mechanisms.

    In the kinetic mode of TNG, the feedback energy is injected as a momentum boost to the gas cells within $\sim2.2$ kpc  at $z=0$ \citep{Zinger2020}. However, as shown in \cite{Weinberger2017}, each energy injection event has a random direction. The injection events occur when the accumulated kinetic feedback energy exceeds a given threshold, thus the gas within feedback region can be accelerated to several tens of thousands of $\kms$. But a coherent gas outflow cannot be built up with the frequent change of energy injection direction. As a result, the kinetic energy is quickly dissipated into thermal energy and heats up the inner gas through shocks within $\sim1$~Myr \citep{Weinberger2017}. The \hi\ density is then significantly lower approaching the centers of quenched galaxies as in Figure~\ref{fig:profile}. The trend with H$_2$ is also quite similar.

    Unlike TNG, SIMBA adopts bipolar injection of the kinetic feedback, i.e. the gas elements surrounding the black holes are ejected parallel to the angular momentum of the inner disks ($\sim256$ nearest gas elements). Although the full jet speed can reach $8000\kms$, the collimated wind outflow only affect a small region of $\le1$~kpc \citep{Dave2019}, which then reduces a small fraction of cold gas in the galactic centre. Since the injection direction is typically stable over the galaxy  dynamical timescale of tens to hundreds of Myrs, the consistent jet energy input will sphericalize at large scales via the hydrodynamical interactions with the CGM gas and keep the halo gas hot. It thus explains the different behaviors of the \hi\ density profiles in SIMBA, with respect to TNG. The QGs in SIMBA still have similar \hi\ densities as the SFGs in the inner regions, while their \hi\ densities are significantly reduced at the outer. As shown in \cite{Appleby2020}, if we set a cold gas density threshold for efficient star formation, SFGs in SIMBA would apparently have more spatially-extended star formation than QGs.  

    When we consider the role that AGN feedback plays in the relation between SFR and \hi\ gas, the weak dependence of the SFR--$M_{\rm HI,2R_{\rm half}}$ relation on $M_{\rm BH}$ and $f_{\rm edd}$ in Figures~\ref{fig:hims_bh} and~\ref{fig:h2ms_bh} does not necessarily mean that AGN feedback is not driving galaxy quenching in TNG and SIMBA. It just reflects that the global star formation law, i.e. the dependence of SFR on \hi\ reservoir, is not significantly affected by AGN feedback. The quenching from AGN feedback is working by depleting the \hi, which thereby reduces the star formation. The effect of AGN feedback will then not show up in the SFR-$M_{\rm HI}$ relation, as long as the instantaneous energy release is not very high and the AGN activity time-scale is much smaller than that of star formation \citep{Guo2022}. It explains the lack of strong evidence in the \hi\ and H$_2$ masses of AGN and non-AGN hosts in observations \citep{Fabello2011,Gereb2015,Ellison2019,Saintonge2017,Shangguan2020}.

    In both simulations, SFGs are dominated by the thermal or radiative AGN modes, while QGs are mostly in the kinetic or jet modes. The kinetic mode is operating with high ${M}_{\rm BH}$ and low $\dot{M}_{\rm BH}$, i.e. low $f_{\rm edd}$. Thus, the Eddington ratio $f_{\rm edd}$ is an indicator for the instantaneous AGN luminosity $L_{\rm AGN}$ at a given $M_{\rm BH}$, since $L_{\rm AGN}$ is proportional to $\dot{M}_{\rm BH}$. The black hole mass $M_{\rm BH}$ represents the cumulative energy release into the surrounding gas, as it is the time integral of $\dot{M}_{\rm BH}$. Our results confirm that the cumulative energy released from AGN activities rather than the instantaneous feedback would reduce the cold gas reservoir and quench the galaxies. It is consistent with the previous finding that quenched massive galaxies are associated with higher integrated energy output from the black holes, while the star-forming and gas-rich galaxies more likely host high-$f_{\rm edd}$ AGNs \citep{Ward2022}. 

	\section{Conclusions}\label{sec:conclusion}
	In this paper, we investigate the cold gas content of star-forming and quenched central galaxies at $z=0$ in three hydrodynamical simulations, Illustris, TNG and SIMBA. By comparing simulations with the observed stacked \hi\ masses of G21 and H$_2$ masses from xCOLD GASS, we find that the observed cold gas properties can be used to effectively constrain the AGN feedback models in simulations, with SIMBA showing the best agreement with observations. We conclude that the cumulative AGN feedback (as traced by black hole mass) is the main force driving the cold gas depletion and thus causes galaxy quenching, but the detailed quenching mechanisms vary for different simulations. Our results are summarized as follows.
	
	(i) Illustris generally does not agree with the observational $M_{\rm HI}$--$M_\ast$ relation. It predicts a weak dependence of $M_{\rm HI}$ on $M_\ast$ for SFGs and significantly under-predicts $M_{\rm HI}$ for QGs, which is caused by its less effective quasar-mode AGN feedback and too efficient radio-mode feedback. 
	
	(ii) TNG improves upon Illustris by increasing the coupling efficiency in the quasar mode and replacing the radio mode with the kinetic feedback, thus showing better agreement with the \hi\ observations. However, it still over-predicts $M_{\rm HI}$ for both the massive star-forming and quenched galaxies. This is clearly shown in the SFR--$M_{\rm HI}$ relation (Fig.~\ref{fig:deltaHI}), where the total $M_{\rm HI}$ only marginally decreases by less than 0.2~dex as the SFR decreases. But as seen from the \hi\ density distributions (Fig.~\ref{fig:profile}), the overall effect of kinetic feedback in TNG is to redistribute the cold gas from the inner regions to the outer. 
	
	(iii) SIMBA agrees best with the \hi\ and H$_2$ observations, by using a two-mode black hole accretion model and a combined kinetic and X-ray feedback mechanism. Unlike the case in TNG, the \hi\ density distribution smoothly increases toward the galaxy centers for both SFGs and QGs, with QGs having significantly reduced cold gas masses in the CGM.  
	
	(iv) When we only consider the cold gas masses within the stellar radii (i.e. $M_{\rm HI,2R_{\rm half}}$ and $M_{\rm H_2,2R_{\rm half}}$), both TNG and SIMBA have very similar decreasing trends of cold gas masses with reduced SFRs (Figs.~\ref{fig:hims_bh} and~\ref{fig:h2ms_bh}), also in good agreement with observations. More importantly, the global star formation law of $\Delta\log{\rm SFR}$--$\Delta\log M_{\rm HI}$ (or $\Delta\log M_{\rm H_2}$) is not significantly affected by AGN feedback. By comparing the effects of $M_{\rm BH}$ and $f_{\rm edd}$, we find that the galaxy quenching is generally achieved by the gradual depletion of cold gas (in the centers for TNG and CGM for SIMBA) due to the cumulative energy release from the AGN activities. But there is an apparent feature of instantaneous cold gas depletion for SFGs in the high-accretion state (i.e. high $f_{\rm edd}$) in the TNG model, likely caused by the efficient quasar-mode feedback.
	
	(v) We measure $\Sigma_1$ in the high-resolution TNG50-1 simulation, and find a very tight correlation between $\Sigma_1$ and $M_{\rm BH}$, with an average scatter of 0.08~dex for $M_{\rm BH}>10^{7.2}\msun$. We also find that the relations of $\Delta\log M_{\rm HI,2R_{\rm half}}$--$\Delta\log\Sigma_1$ and $\Delta\log M_{\rm H_2,2R_{\rm half}}$--$\Delta\log\Sigma_1$ from TNG50 are in very good agreement with those from observations. It suggests that the observed decreasing trend of the cold gas masses with increasing $\Sigma_1$ (G21) is very likely driven by the growth of central black holes.
	
	In conclusion, the masses of cold gas, as well as its distribution in galaxies, provide strong constraints on AGN feedback models in the current hydrodynamical simulations. In the three simulation models investigated in this paper, galaxy quenching is caused by the combined effects of cumulative AGN feedback from the central black hole growth and the subgrid models of star formation. 
	
	The kinetic feedback mechanism in TNG is too weak to repel enough cold gas far away from the central galaxies, causing large discrepancies with the cold gas observations. But the feedback models in SIMBA are not necessarily correct, as seen from some discrepancies with other observations \citep{Dave2020}. Future \hi\ and H$_2$ surveys, e.g. WALLABY \citep{Koribalski2020}, Apertif \citep{Verheijen2008} and CRAFTS \citep{Li2018}, at higher redshifts and over larger volumes will provide an excellent data set to examine these AGN feedback models.
	
	\begin{acknowledgments}
	    We thank the anonymous reviewer for the helpful comments that significantly improved the presentation of this paper. We thank Zhu Chen and Lin Lin for helpful discussions. 
		This work is supported by the National Key R\&D Program of China (grant No. 2018YFA0404503), National SKA Program of China (grant No. 2020SKA0110100), National Science Foundation of China (Nos. 11922305, 11833005, 12073002, 11721303, 12011530159) and the science research grants from the China Manned Space Project with NOs. CMS-CSST-2021-A01, CMS-CSST-2021-A02 and CMS-CSST-2021-B01. W.C. is supported by the STFC AGP Grant ST/V000594/1 and the Atracci\'{o}n de Talento Contract no. 2020-T1/TIC-19882 granted by the Comunidad de Madrid in Spain. We acknowledge the use of the High Performance Computing Resource in the Core Facility for Advanced Research Computing at the Shanghai Astronomical Observatory.
	\end{acknowledgments}
	
	\appendix
	\section{H{~\scriptsize I} $/ \rm H_2$ transition models in TNG}\label{app}
	\begin{figure*}
		\centering
		\includegraphics[width=0.75\textwidth]{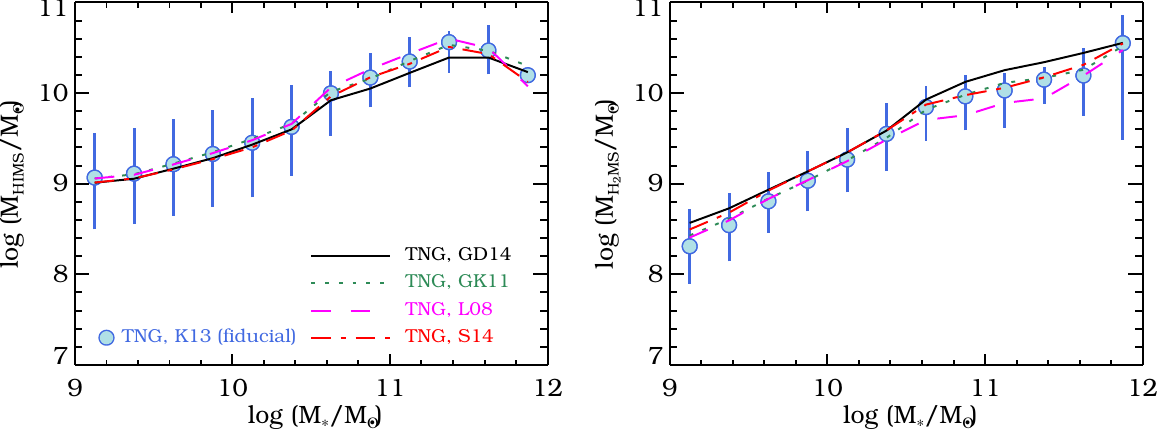}
		\caption{The trends of HIMS (left panel) and H$_2$MS (right panel) for five \hi $/ H_2$ transition models in TNG. Compared to Figure~\ref{fig:hims}, we show K13 model, which we used in this paper, as filled circles as fiducial data. The black solid line, green dotted line, purple dashed line and red dashdotted line represent GD14, GK11, L08 and S14 models, respectively.}
		\label{fig:model_ms}
	\end{figure*}
	\begin{figure*}
		\centering
		\includegraphics[width=0.75\textwidth]{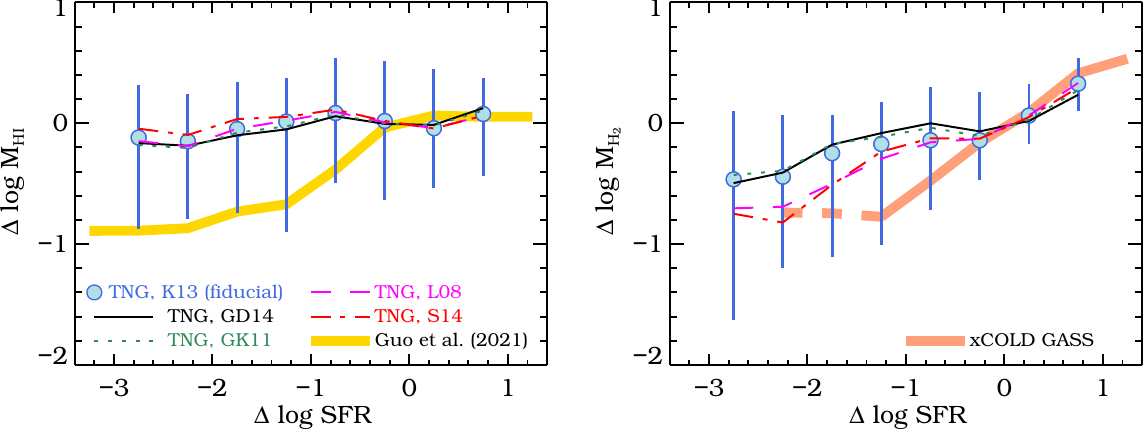}
		\caption{Relation between $\rm {\Delta \log SFR}$ and $\rm {\Delta \log M_{HI}}$ (left panel), $\rm {\Delta \log M_{H_2}}$ (right panel) for five \hi $/ H_2$ transition models in TNG. Labels are the same as Figure~\ref{fig:model_ms}. Similar to Figure~\ref{fig:deltaHI}, we show G21 and xCOLD GASS observational data for comparison.}
		\label{fig:model}
	\end{figure*}
	\begin{figure*}
		\centering
		\includegraphics[width=0.65\textwidth]{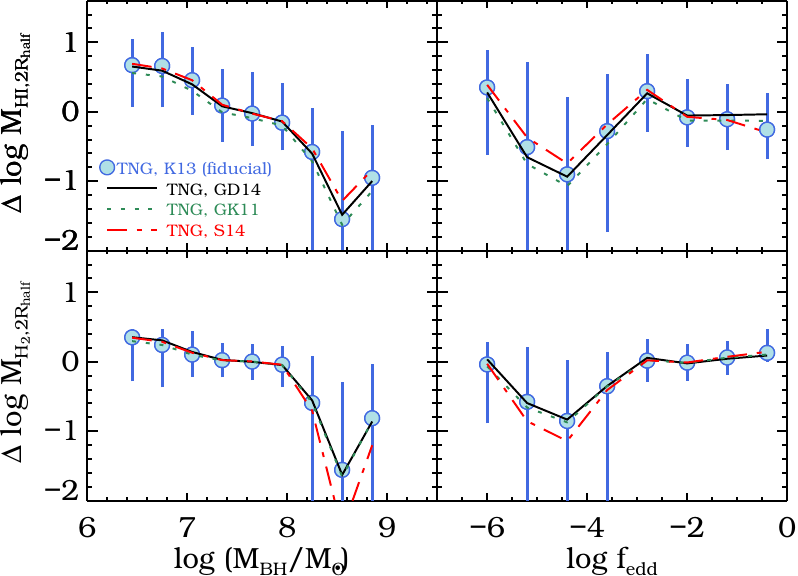}
		\caption{Similar to Figure~\ref{fig:himsdisk}, we show the dependence of $\rm {\Delta \log M_{HI, 2R_{half}}}$ (top panels), $\rm {\Delta \log M_{H_2, 2R_{half}}}$ (bottom panels) on $M_{\rm BH}$ (left panels) and $f_{\rm edd}$ (right panels) for five \hi $/ H_2$ transition models in TNG. Labels are the same as Figure~\ref{fig:model_ms}.}
		\label{fig:model_disk}
	\end{figure*}
	
TNG and Illustris adopt five different \hi\ to $\rm H_2$ transition models, i.e. \citet[L08]{Leroy2008}, \citet[GK11]{Gnedin2011}, \citet[K13]{Krumholz2013}, \citet[GD14]{Gnedin2014} and \citet[S14]{Sternberg2014}. To investigate the effects of different models on our results, we show the comparisons on definitions of HIMS and H$_2$MS (Figure~\ref{fig:model_ms}), relation between $\Delta\log M_{\rm HI}$ and $\Delta\log{\rm SFR}$ (Figure~\ref{fig:model}) and dependences of \hi\ and H$_2$ gas within $2R_{\rm half}$ on $M_{\rm BH}$ and $f_{\rm edd}$ (Figure~\ref{fig:model_disk}). The L08 model is not shown in Figure~\ref{fig:model_disk} due to the lack of \hi\ and H$_2$ density profiles in the public release of TNG. 

As we see, there is no strong effect of different transition models on our results, especially for the \hi\ gas. The influence of different models is slightly stronger for H$_2$ shown in Figure~\ref{fig:model}, with L08 and S14 models predicting slightly lower $M_{\rm H2}$ for QGs.  Thus, our conclusion of the quenching mechanisms in TNG is not affected by the different transition models. It is consistent with our previous discussion that neither \hi\ or H$_2$ is directly related to the change of SFR in TNG, but rather the total neutral hydrogen density.
    
    \bibliographystyle{aasjournal}

\begin{thebibliography}{}
	\expandafter\ifx\csname natexlab\endcsname\relax\def\natexlab#1{#1}\fi
	\providecommand{\url}[1]{\href{#1}{#1}}
	\providecommand{\dodoi}[1]{doi:~\href{http://doi.org/#1}{\nolinkurl{#1}}}
	\providecommand{\doeprint}[1]{\href{http://ascl.net/#1}{\nolinkurl{http://ascl.net/#1}}}
	\providecommand{\doarXiv}[1]{\href{https://arxiv.org/abs/#1}{\nolinkurl{https://arxiv.org/abs/#1}}}
	
	\bibitem[{{Abazajian} {et~al.}(2009){Abazajian}, {Adelman-McCarthy},
		{Ag{\"u}eros}, {Allam}, {Allende Prieto}, {An}, {Anderson}, {Anderson},
		{Annis}, {Bahcall}, {Bailer-Jones}, {Barentine}, {Bassett}, {Becker},
		{Beers}, {Bell}, {Belokurov}, {Berlind}, {Berman}, {Bernardi}, {Bickerton},
		{Bizyaev}, {Blakeslee}, {Blanton}, {Bochanski}, {Boroski}, {Brewington},
		{Brinchmann}, {Brinkmann}, {Brunner}, {Budav{\'a}ri}, {Carey}, {Carliles},
		{Carr}, {Castander}, {Cinabro}, {Connolly}, {Csabai}, {Cunha}, {Czarapata},
		{Davenport}, {de Haas}, {Dilday}, {Doi}, {Eisenstein}, {Evans}, {Evans},
		{Fan}, {Friedman}, {Frieman}, {Fukugita}, {G{\"a}nsicke}, {Gates},
		{Gillespie}, {Gilmore}, {Gonzalez}, {Gonzalez}, {Grebel}, {Gunn},
		{Gy{\"o}ry}, {Hall}, {Harding}, {Harris}, {Harvanek}, {Hawley}, {Hayes},
		{Heckman}, {Hendry}, {Hennessy}, {Hindsley}, {Hoblitt}, {Hogan}, {Hogg},
		{Holtzman}, {Hyde}, {Ichikawa}, {Ichikawa}, {Im}, {Ivezi{\'c}}, {Jester},
		{Jiang}, {Johnson}, {Jorgensen}, {Juri{\'c}}, {Kent}, {Kessler}, {Kleinman},
		{Knapp}, {Konishi}, {Kron}, {Krzesinski}, {Kuropatkin}, {Lampeitl},
		{Lebedeva}, {Lee}, {Lee}, {French Leger}, {L{\'e}pine}, {Li}, {Lima}, {Lin},
		{Long}, {Loomis}, {Loveday}, {Lupton}, {Magnier}, {Malanushenko},
		{Malanushenko}, {Mandelbaum}, {Margon}, {Marriner}, {Mart{\'\i}nez-Delgado},
		{Matsubara}, {McGehee}, {McKay}, {Meiksin}, {Morrison}, {Mullally}, {Munn},
		{Murphy}, {Nash}, {Nebot}, {Neilsen}, {Newberg}, {Newman}, {Nichol},
		{Nicinski}, {Nieto-Santisteban}, {Nitta}, {Okamura}, {Oravetz}, {Ostriker},
		{Owen}, {Padmanabhan}, {Pan}, {Park}, {Pauls}, {Peoples}, {Percival}, {Pier},
		{Pope}, {Pourbaix}, {Price}, {Purger}, {Quinn}, {Raddick}, {Re Fiorentin},
		{Richards}, {Richmond}, {Riess}, {Rix}, {Rockosi}, {Sako}, {Schlegel},
		{Schneider}, {Scholz}, {Schreiber}, {Schwope}, {Seljak}, {Sesar}, {Sheldon},
		{Shimasaku}, {Sibley}, {Simmons}, {Sivarani}, {Allyn Smith}, {Smith},
		{Smol{\v{c}}i{\'c}}, {Snedden}, {Stebbins}, {Steinmetz}, {Stoughton},
		{Strauss}, {SubbaRao}, {Suto}, {Szalay}, {Szapudi}, {Szkody}, {Tanaka},
		{Tegmark}, {Teodoro}, {Thakar}, {Tremonti}, {Tucker}, {Uomoto}, {Vanden
			Berk}, {Vandenberg}, {Vidrih}, {Vogeley}, {Voges}, {Vogt}, {Wadadekar},
		{Watters}, {Weinberg}, {West}, {White}, {Wilhite}, {Wonders}, {Yanny},
		{Yocum}, {York}, {Zehavi}, {Zibetti}, \& {Zucker}}]{Abazajian2009}
	{Abazajian}, K.~N., {Adelman-McCarthy}, J.~K., {Ag{\"u}eros}, M.~A., {et~al.}
	2009, \apjs, 182, 543, \dodoi{10.1088/0067-0049/182/2/543}
	
	\bibitem[{{Angl{\'e}s-Alc{\'a}zar} {et~al.}(2017){Angl{\'e}s-Alc{\'a}zar},
		{Dav{\'e}}, {Faucher-Gigu{\`e}re}, {{\"O}zel}, \& {Hopkins}}]{Angles2017}
	{Angl{\'e}s-Alc{\'a}zar}, D., {Dav{\'e}}, R., {Faucher-Gigu{\`e}re}, C.-A.,
	{{\"O}zel}, F., \& {Hopkins}, P.~F. 2017, \mnras, 464, 2840,
	\dodoi{10.1093/mnras/stw2565}
	
	\bibitem[{{Angl{\'e}s-Alc{\'a}zar} {et~al.}(2013){Angl{\'e}s-Alc{\'a}zar},
		{{\"O}zel}, \& {Dav{\'e}}}]{Angles2013}
	{Angl{\'e}s-Alc{\'a}zar}, D., {{\"O}zel}, F., \& {Dav{\'e}}, R. 2013, \apj,
	770, 5, \dodoi{10.1088/0004-637X/770/1/5}
	
	\bibitem[{{Angl{\'e}s-Alc{\'a}zar} {et~al.}(2015){Angl{\'e}s-Alc{\'a}zar},
		{{\"O}zel}, {Dav{\'e}}, {Katz}, {Kollmeier}, \& {Oppenheimer}}]{Angles2015}
	{Angl{\'e}s-Alc{\'a}zar}, D., {{\"O}zel}, F., {Dav{\'e}}, R., {et~al.} 2015,
	\apj, 800, 127, \dodoi{10.1088/0004-637X/800/2/127}
	
	\bibitem[{{Appleby} {et~al.}(2020){Appleby}, {Dav{\'e}}, {Kraljic},
		{Angl{\'e}s-Alc{\'a}zar}, \& {Narayanan}}]{Appleby2020}
	{Appleby}, S., {Dav{\'e}}, R., {Kraljic}, K., {Angl{\'e}s-Alc{\'a}zar}, D., \&
	{Narayanan}, D. 2020, \mnras, 494, 6053, \dodoi{10.1093/mnras/staa1169}
	
	\bibitem[{{Barro} {et~al.}(2017){Barro}, {Faber}, {Koo}, {Dekel}, {Fang},
		{Trump}, {P{\'e}rez-Gonz{\'a}lez}, {Pacifici}, {Primack}, {Somerville},
		{Yan}, {Guo}, {Liu}, {Ceverino}, {Kocevski}, \& {McGrath}}]{Barro2017}
	{Barro}, G., {Faber}, S.~M., {Koo}, D.~C., {et~al.} 2017, \apj, 840, 47,
	\dodoi{10.3847/1538-4357/aa6b05}
	
	\bibitem[{{Beckmann} {et~al.}(2017){Beckmann}, {Devriendt}, {Slyz}, {Peirani},
		{Richardson}, {Dubois}, {Pichon}, {Chisari}, {Kaviraj}, {Laigle}, \&
		{Volonteri}}]{Beckmann2017}
	{Beckmann}, R.~S., {Devriendt}, J., {Slyz}, A., {et~al.} 2017, \mnras, 472,
	949, \dodoi{10.1093/mnras/stx1831}
	
	\bibitem[{{Bigiel} {et~al.}(2008){Bigiel}, {Leroy}, {Walter}, {Brinks}, {de
			Blok}, {Madore}, \& {Thornley}}]{Bigiel2008}
	{Bigiel}, F., {Leroy}, A., {Walter}, F., {et~al.} 2008, \aj, 136, 2846,
	\dodoi{10.1088/0004-6256/136/6/2846}
	
	\bibitem[{{Birnboim} \& {Dekel}(2003)}]{Birnboim2003}
	{Birnboim}, Y., \& {Dekel}, A. 2003, \mnras, 345, 349,
	\dodoi{10.1046/j.1365-8711.2003.06955.x}
	
	\bibitem[{{Bower} {et~al.}(2006){Bower}, {Benson}, {Malbon}, {Helly}, {Frenk},
		{Baugh}, {Cole}, \& {Lacey}}]{Bower2006}
	{Bower}, R.~G., {Benson}, A.~J., {Malbon}, R., {et~al.} 2006, \mnras, 370, 645,
	\dodoi{10.1111/j.1365-2966.2006.10519.x}
	
	\bibitem[{{Brown} {et~al.}(2017){Brown}, {Catinella}, {Cortese}, {Lagos},
		{Dav{\'e}}, {Kilborn}, {Haynes}, {Giovanelli}, \&
		{Rafieferantsoa}}]{Brown2017}
	{Brown}, T., {Catinella}, B., {Cortese}, L., {et~al.} 2017, \mnras, 466, 1275,
	\dodoi{10.1093/mnras/stw2991}
	
	\bibitem[{{Catinella} {et~al.}(2010){Catinella}, {Schiminovich}, {Kauffmann},
		{Fabello}, {Wang}, {Hummels}, {Lemonias}, {Moran}, {Wu}, {Giovanelli},
		{Haynes}, {Heckman}, {Basu-Zych}, {Blanton}, {Brinchmann}, {Budav{\'a}ri},
		{Gon{\c{c}}alves}, {Johnson}, {Kennicutt}, {Madore}, {Martin}, {Rich},
		{Tacconi}, {Thilker}, {Wild}, \& {Wyder}}]{Catinella2010}
	{Catinella}, B., {Schiminovich}, D., {Kauffmann}, G., {et~al.} 2010, \mnras,
	403, 683, \dodoi{10.1111/j.1365-2966.2009.16180.x}
	
	\bibitem[{{Catinella} {et~al.}(2018){Catinella}, {Saintonge}, {Janowiecki},
		{Cortese}, {Dav{\'e}}, {Lemonias}, {Cooper}, {Schiminovich}, {Hummels},
		{Fabello}, {Ger{\'e}b}, {Kilborn}, \& {Wang}}]{Catinella2018}
	{Catinella}, B., {Saintonge}, A., {Janowiecki}, S., {et~al.} 2018, \mnras, 476,
	875, \dodoi{10.1093/mnras/sty089}
	
	\bibitem[{{Chauhan} {et~al.}(2021){Chauhan}, {Lagos}, {Stevens}, {Bravo},
		{Rhee}, {Power}, {Obreschkow}, \& {Meyer}}]{Chauhan2021}
	{Chauhan}, G., {Lagos}, C. d.~P., {Stevens}, A. R.~H., {et~al.} 2021, \mnras,
	506, 4893, \dodoi{10.1093/mnras/stab1925}
	
	\bibitem[{{Chen} {et~al.}(2020){Chen}, {Faber}, {Koo}, {Somerville}, {Primack},
		{Dekel}, {Rodr{\'\i}guez-Puebla}, {Guo}, {Barro}, {Kocevski}, {van der Wel},
		{Woo}, {Bell}, {Fang}, {Ferguson}, {Giavalisco}, {Huertas-Company}, {Jiang},
		{Kassin}, {Lin}, {Liu}, {Luo}, {Luo}, {Pacifici}, {Pandya}, {Salim}, {Shu},
		{Tacchella}, {Terrazas}, \& {Yesuf}}]{Chen2020}
	{Chen}, Z., {Faber}, S.~M., {Koo}, D.~C., {et~al.} 2020, \apj, 897, 102,
	\dodoi{10.3847/1538-4357/ab9633}
	
	\bibitem[{{Cheung} {et~al.}(2012){Cheung}, {Faber}, {Koo}, {Dutton}, {Simard},
		{McGrath}, {Huang}, {Bell}, {Dekel}, {Fang}, {Salim}, {Barro}, {Bundy},
		{Coil}, {Cooper}, {Conselice}, {Davis}, {Dom{\'\i}nguez}, {Kassin},
		{Kocevski}, {Koekemoer}, {Lin}, {Lotz}, {Newman}, {Phillips}, {Rosario},
		{Weiner}, \& {Willmer}}]{Cheung2012}
	{Cheung}, E., {Faber}, S.~M., {Koo}, D.~C., {et~al.} 2012, \apj, 760, 131,
	\dodoi{10.1088/0004-637X/760/2/131}
	
	\bibitem[{{Cortese} {et~al.}(2021){Cortese}, {Catinella}, \&
		{Smith}}]{Cortese2021}
	{Cortese}, L., {Catinella}, B., \& {Smith}, R. 2021, \pasa, 38, e035,
	\dodoi{10.1017/pasa.2021.18}
	
	\bibitem[{{Crain} {et~al.}(2015){Crain}, {Schaye}, {Bower}, {Furlong},
		{Schaller}, {Theuns}, {Dalla Vecchia}, {Frenk}, {McCarthy}, {Helly},
		{Jenkins}, {Rosas-Guevara}, {White}, \& {Trayford}}]{Crain2015}
	{Crain}, R.~A., {Schaye}, J., {Bower}, R.~G., {et~al.} 2015, \mnras, 450, 1937,
	\dodoi{10.1093/mnras/stv725}
	
	\bibitem[{{Croton} {et~al.}(2006){Croton}, {Springel}, {White}, {De Lucia},
		{Frenk}, {Gao}, {Jenkins}, {Kauffmann}, {Navarro}, \& {Yoshida}}]{Croton2006}
	{Croton}, D.~J., {Springel}, V., {White}, S. D.~M., {et~al.} 2006, \mnras, 365,
	11, \dodoi{10.1111/j.1365-2966.2005.09675.x}
	
	\bibitem[{{Cui} {et~al.}(2021){Cui}, {Dav{\'e}}, {Peacock},
		{Angl{\'e}s-Alc{\'a}zar}, \& {Yang}}]{Cui2021}
	{Cui}, W., {Dav{\'e}}, R., {Peacock}, J.~A., {Angl{\'e}s-Alc{\'a}zar}, D., \&
	{Yang}, X. 2021, Nature Astronomy, 5, 1069,
	\dodoi{10.1038/s41550-021-01404-1}
	
	\bibitem[{{Dav{\'e}} {et~al.}(2019){Dav{\'e}}, {Angl{\'e}s-Alc{\'a}zar},
		{Narayanan}, {Li}, {Rafieferantsoa}, \& {Appleby}}]{Dave2019}
	{Dav{\'e}}, R., {Angl{\'e}s-Alc{\'a}zar}, D., {Narayanan}, D., {et~al.} 2019,
	\mnras, 486, 2827, \dodoi{10.1093/mnras/stz937}
	
	\bibitem[{{Dav{\'e}} {et~al.}(2020){Dav{\'e}}, {Crain}, {Stevens}, {Narayanan},
		{Saintonge}, {Catinella}, \& {Cortese}}]{Dave2020}
	{Dav{\'e}}, R., {Crain}, R.~A., {Stevens}, A. R.~H., {et~al.} 2020, \mnras,
	497, 146, \dodoi{10.1093/mnras/staa1894}
	
	\bibitem[{{Dav{\'e}} {et~al.}(2016){Dav{\'e}}, {Thompson}, \&
		{Hopkins}}]{Dave2016}
	{Dav{\'e}}, R., {Thompson}, R., \& {Hopkins}, P.~F. 2016, \mnras, 462, 3265,
	\dodoi{10.1093/mnras/stw1862}
	
	\bibitem[{{Davies} {et~al.}(2020){Davies}, {Crain}, {Oppenheimer}, \&
		{Schaye}}]{Davies2020}
	{Davies}, J.~J., {Crain}, R.~A., {Oppenheimer}, B.~D., \& {Schaye}, J. 2020,
	\mnras, 491, 4462, \dodoi{10.1093/mnras/stz3201}
	
	\bibitem[{{Dekel} \& {Birnboim}(2006)}]{Dekel2006}
	{Dekel}, A., \& {Birnboim}, Y. 2006, \mnras, 368, 2,
	\dodoi{10.1111/j.1365-2966.2006.10145.x}
	
	\bibitem[{{Dekel} \& {Burkert}(2014)}]{Dekel2014}
	{Dekel}, A., \& {Burkert}, A. 2014, \mnras, 438, 1870,
	\dodoi{10.1093/mnras/stt2331}
	
	\bibitem[{{Di Matteo} {et~al.}(2005){Di Matteo}, {Springel}, \&
		{Hernquist}}]{Matteo2005}
	{Di Matteo}, T., {Springel}, V., \& {Hernquist}, L. 2005, \nat, 433, 604,
	\dodoi{10.1038/nature03335}
	
	\bibitem[{{Diemer} {et~al.}(2018){Diemer}, {Stevens}, {Forbes}, {Marinacci},
		{Hernquist}, {Lagos}, {Sternberg}, {Pillepich}, {Nelson}, {Popping},
		{Villaescusa-Navarro}, {Torrey}, \& {Vogelsberger}}]{Diemer2018}
	{Diemer}, B., {Stevens}, A. R.~H., {Forbes}, J.~C., {et~al.} 2018, \apjs, 238,
	33, \dodoi{10.3847/1538-4365/aae387}
	
	\bibitem[{{Diemer} {et~al.}(2019){Diemer}, {Stevens}, {Lagos}, {Calette},
		{Tacchella}, {Hernquist}, {Marinacci}, {Nelson}, {Pillepich},
		{Rodriguez-Gomez}, {Villaescusa-Navarro}, \& {Vogelsberger}}]{Diemer2019}
	{Diemer}, B., {Stevens}, A. R.~H., {Lagos}, C. d.~P., {et~al.} 2019, \mnras,
	487, 1529, \dodoi{10.1093/mnras/stz1323}
	
	\bibitem[{{Donnari} {et~al.}(2021){Donnari}, {Pillepich}, {Nelson},
		{Marinacci}, {Vogelsberger}, \& {Hernquist}}]{Donnari2021}
	{Donnari}, M., {Pillepich}, A., {Nelson}, D., {et~al.} 2021, \mnras, 506, 4760,
	\dodoi{10.1093/mnras/stab1950}
	
	\bibitem[{{Donnari} {et~al.}(2019){Donnari}, {Pillepich}, {Nelson},
		{Vogelsberger}, {Genel}, {Weinberger}, {Marinacci}, {Springel}, \&
		{Hernquist}}]{Donnari2019}
	---. 2019, \mnras, 485, 4817, \dodoi{10.1093/mnras/stz712}
	
	\bibitem[{{Dubois} {et~al.}(2016){Dubois}, {Peirani}, {Pichon}, {Devriendt},
		{Gavazzi}, {Welker}, \& {Volonteri}}]{Dubois2016}
	{Dubois}, Y., {Peirani}, S., {Pichon}, C., {et~al.} 2016, \mnras, 463, 3948,
	\dodoi{10.1093/mnras/stw2265}
	
	\bibitem[{{Ellison} {et~al.}(2019){Ellison}, {Brown}, {Catinella}, \&
		{Cortese}}]{Ellison2019}
	{Ellison}, S.~L., {Brown}, T., {Catinella}, B., \& {Cortese}, L. 2019, \mnras,
	482, 5694, \dodoi{10.1093/mnras/sty3139}
	
	\bibitem[{{Ellison} {et~al.}(2021){Ellison}, {Wong}, {S{\'a}nchez}, {Colombo},
		{Bolatto}, {Barrera-Ballesteros}, {Garc{\'\i}a-Benito}, {Kalinova}, {Luo},
		{Rubio}, \& {Vogel}}]{Ellison2021}
	{Ellison}, S.~L., {Wong}, T., {S{\'a}nchez}, S.~F., {et~al.} 2021, \mnras, 505,
	L46, \dodoi{10.1093/mnrasl/slab047}
	
	\bibitem[{{Fabello} {et~al.}(2011){Fabello}, {Kauffmann}, {Catinella},
		{Giovanelli}, {Haynes}, {Heckman}, \& {Schiminovich}}]{Fabello2011}
	{Fabello}, S., {Kauffmann}, G., {Catinella}, B., {et~al.} 2011, \mnras, 416,
	1739, \dodoi{10.1111/j.1365-2966.2011.18825.x}
	
	\bibitem[{{Fang} {et~al.}(2013){Fang}, {Faber}, {Koo}, \& {Dekel}}]{Fang2013}
	{Fang}, J.~J., {Faber}, S.~M., {Koo}, D.~C., \& {Dekel}, A. 2013, \apj, 776,
	63, \dodoi{10.1088/0004-637X/776/1/63}
	
	\bibitem[{{Fletcher} {et~al.}(2021){Fletcher}, {Saintonge}, {Soares}, \&
		{Pontzen}}]{Fletcher2021}
	{Fletcher}, T.~J., {Saintonge}, A., {Soares}, P.~S., \& {Pontzen}, A. 2021,
	\mnras, 501, 411, \dodoi{10.1093/mnras/staa3025}
	
	\bibitem[{{Fluetsch} {et~al.}(2019){Fluetsch}, {Maiolino}, {Carniani},
		{Marconi}, {Cicone}, {Bourne}, {Costa}, {Fabian}, {Ishibashi}, \&
		{Venturi}}]{Fluetsch2019}
	{Fluetsch}, A., {Maiolino}, R., {Carniani}, S., {et~al.} 2019, \mnras, 483,
	4586, \dodoi{10.1093/mnras/sty3449}
	
	\bibitem[{{Genel} {et~al.}(2014){Genel}, {Vogelsberger}, {Springel}, {Sijacki},
		{Nelson}, {Snyder}, {Rodriguez-Gomez}, {Torrey}, \& {Hernquist}}]{Genel2014}
	{Genel}, S., {Vogelsberger}, M., {Springel}, V., {et~al.} 2014, \mnras, 445,
	175, \dodoi{10.1093/mnras/stu1654}
	
	\bibitem[{{Ger{\'e}b} {et~al.}(2015){Ger{\'e}b}, {Morganti}, {Oosterloo},
		{Hoppmann}, \& {Staveley-Smith}}]{Gereb2015}
	{Ger{\'e}b}, K., {Morganti}, R., {Oosterloo}, T.~A., {Hoppmann}, L., \&
	{Staveley-Smith}, L. 2015, \aap, 580, A43,
	\dodoi{10.1051/0004-6361/201424810}
	
	\bibitem[{{Giovanelli} {et~al.}(2005){Giovanelli}, {Haynes}, {Kent},
		{Perillat}, {Saintonge}, {Brosch}, {Catinella}, {Hoffman}, {Stierwalt},
		{Spekkens}, {Lerner}, {Masters}, {Momjian}, {Rosenberg}, {Springob},
		{Boselli}, {Charmandaris}, {Darling}, {Davies}, {Garcia Lambas}, {Gavazzi},
		{Giovanardi}, {Hardy}, {Hunt}, {Iovino}, {Karachentsev}, {Karachentseva},
		{Koopmann}, {Marinoni}, {Minchin}, {Muller}, {Putman}, {Pantoja}, {Salzer},
		{Scodeggio}, {Skillman}, {Solanes}, {Valotto}, {van Driel}, \& {van
			Zee}}]{Giovanelli2005}
	{Giovanelli}, R., {Haynes}, M.~P., {Kent}, B.~R., {et~al.} 2005, \aj, 130,
	2598, \dodoi{10.1086/497431}
	
	\bibitem[{{Gnedin} \& {Draine}(2014)}]{Gnedin2014}
	{Gnedin}, N.~Y., \& {Draine}, B.~T. 2014, \apj, 795, 37,
	\dodoi{10.1088/0004-637X/795/1/37}
	
	\bibitem[{{Gnedin} \& {Kravtsov}(2011)}]{Gnedin2011}
	{Gnedin}, N.~Y., \& {Kravtsov}, A.~V. 2011, \apj, 728, 88,
	\dodoi{10.1088/0004-637X/728/2/88}
	
	\bibitem[{{Guo} {et~al.}(2020){Guo}, {Jones}, {Haynes}, \& {Fu}}]{Guo2020}
	{Guo}, H., {Jones}, M.~G., {Haynes}, M.~P., \& {Fu}, J. 2020, \apj, 894, 92,
	\dodoi{10.3847/1538-4357/ab886f}
	
	\bibitem[{{Guo} {et~al.}(2022){Guo}, {Jones}, \& {Wang}}]{Guo2022}
	{Guo}, H., {Jones}, M.~G., \& {Wang}, J. 2022, \apjl, 933, L12,
	\dodoi{10.3847/2041-8213/ac794f}
	
	\bibitem[{{Guo} {et~al.}(2021){Guo}, {Jones}, {Wang}, \& {Lin}}]{Guo2021}
	{Guo}, H., {Jones}, M.~G., {Wang}, J., \& {Lin}, L. 2021, \apj, 918, 53,
	\dodoi{10.3847/1538-4357/ac062e}
	
	\bibitem[{{Guo} {et~al.}(2017){Guo}, {Li}, {Zheng}, {Mo}, {Jing}, {Zu}, {Lim},
		\& {Xu}}]{Guo2017}
	{Guo}, H., {Li}, C., {Zheng}, Z., {et~al.} 2017, \apj, 846, 61,
	\dodoi{10.3847/1538-4357/aa85e7}
	
	\bibitem[{{Habouzit} {et~al.}(2021){Habouzit}, {Li}, {Somerville}, {Genel},
		{Pillepich}, {Volonteri}, {Dav{\'e}}, {Rosas-Guevara}, {McAlpine}, {Peirani},
		{Hernquist}, {Angl{\'e}s-Alc{\'a}zar}, {Reines}, {Bower}, {Dubois}, {Nelson},
		{Pichon}, \& {Vogelsberger}}]{Habouzit2021}
	{Habouzit}, M., {Li}, Y., {Somerville}, R.~S., {et~al.} 2021, \mnras, 503,
	1940, \dodoi{10.1093/mnras/stab496}
	
	\bibitem[{{Hahn} {et~al.}(2019){Hahn}, {Starkenburg}, {Choi}, {Dav{\'e}},
		{Dickey}, {Geha}, {Genel}, {Hayward}, {Maller}, {Mandyam}, {Pandya},
		{Popping}, {Rafieferantsoa}, {Somerville}, \& {Tinker}}]{Hahn2019}
	{Hahn}, C., {Starkenburg}, T.~K., {Choi}, E., {et~al.} 2019, \apj, 872, 160,
	\dodoi{10.3847/1538-4357/aafedd}
	
	\bibitem[{{Haynes} {et~al.}(2011){Haynes}, {Giovanelli}, {Martin}, {Hess},
		{Saintonge}, {Adams}, {Hallenbeck}, {Hoffman}, {Huang}, {Kent}, {Koopmann},
		{Papastergis}, {Stierwalt}, {Balonek}, {Craig}, {Higdon}, {Kornreich},
		{Miller}, {O'Donoghue}, {Olowin}, {Rosenberg}, {Spekkens}, {Troischt}, \&
		{Wilcots}}]{Haynes2011}
	{Haynes}, M.~P., {Giovanelli}, R., {Martin}, A.~M., {et~al.} 2011, \aj, 142,
	170, \dodoi{10.1088/0004-6256/142/5/170}
	
	\bibitem[{{Haynes} {et~al.}(2018){Haynes}, {Giovanelli}, {Kent}, {Adams},
		{Balonek}, {Craig}, {Fertig}, {Finn}, {Giovanardi}, {Hallenbeck}, {Hess},
		{Hoffman}, {Huang}, {Jones}, {Koopmann}, {Kornreich}, {Leisman}, {Miller},
		{Moorman}, {O'Connor}, {O'Donoghue}, {Papastergis}, {Troischt}, {Stark}, \&
		{Xiao}}]{Haynes2018}
	{Haynes}, M.~P., {Giovanelli}, R., {Kent}, B.~R., {et~al.} 2018, \apj, 861, 49,
	\dodoi{10.3847/1538-4357/aac956}
	
	\bibitem[{{Heckman} \& {Best}(2014)}]{Heckman2014}
	{Heckman}, T.~M., \& {Best}, P.~N. 2014, \araa, 52, 589,
	\dodoi{10.1146/annurev-astro-081913-035722}
	
	\bibitem[{{Hopkins}(2015)}]{Hopkins2015}
	{Hopkins}, P.~F. 2015, \mnras, 450, 53, \dodoi{10.1093/mnras/stv195}
	
	\bibitem[{{Hopkins} \& {Quataert}(2011)}]{Hopkins2011}
	{Hopkins}, P.~F., \& {Quataert}, E. 2011, \mnras, 415, 1027,
	\dodoi{10.1111/j.1365-2966.2011.18542.x}
	
	\bibitem[{{Ishibashi} \& {Fabian}(2012)}]{Ishibashi2012}
	{Ishibashi}, W., \& {Fabian}, A.~C. 2012, \mnras, 427, 2998,
	\dodoi{10.1111/j.1365-2966.2012.22074.x}
	
	\bibitem[{{Janowiecki} {et~al.}(2020){Janowiecki}, {Catinella}, {Cortese},
		{Saintonge}, \& {Wang}}]{Janowiecki2020}
	{Janowiecki}, S., {Catinella}, B., {Cortese}, L., {Saintonge}, A., \& {Wang},
	J. 2020, \mnras, 493, 1982, \dodoi{10.1093/mnras/staa178}
	
	\bibitem[{{Jones} {et~al.}(2018){Jones}, {Haynes}, {Giovanelli}, \&
		{Moorman}}]{Jones2018}
	{Jones}, M.~G., {Haynes}, M.~P., {Giovanelli}, R., \& {Moorman}, C. 2018,
	\mnras, 477, 2, \dodoi{10.1093/mnras/sty521}
	
	\bibitem[{{Kaviraj} {et~al.}(2017){Kaviraj}, {Laigle}, {Kimm}, {Devriendt},
		{Dubois}, {Pichon}, {Slyz}, {Chisari}, \& {Peirani}}]{Kaviraj2017}
	{Kaviraj}, S., {Laigle}, C., {Kimm}, T., {et~al.} 2017, \mnras, 467, 4739,
	\dodoi{10.1093/mnras/stx126}
	
	\bibitem[{{Kennicutt}(1998)}]{Kennicutt1998}
	{Kennicutt}, Robert~C., J. 1998, \apj, 498, 541, \dodoi{10.1086/305588}
	
	\bibitem[{{Kennicutt} {et~al.}(2007){Kennicutt}, {Calzetti}, {Walter}, {Helou},
		{Hollenbach}, {Armus}, {Bendo}, {Dale}, {Draine}, {Engelbracht}, {Gordon},
		{Prescott}, {Regan}, {Thornley}, {Bot}, {Brinks}, {de Blok}, {de Mello},
		{Meyer}, {Moustakas}, {Murphy}, {Sheth}, \& {Smith}}]{Kennicutt2007}
	{Kennicutt}, Robert~C., J., {Calzetti}, D., {Walter}, F., {et~al.} 2007, \apj,
	671, 333, \dodoi{10.1086/522300}
	
	\bibitem[{{Keres} {et~al.}(2003){Keres}, {Yun}, \& {Young}}]{Keres2003}
	{Keres}, D., {Yun}, M.~S., \& {Young}, J.~S. 2003, \apj, 582, 659,
	\dodoi{10.1086/344820}
	
	\bibitem[{{Koribalski} {et~al.}(2020){Koribalski}, {Staveley-Smith},
		{Westmeier}, {Serra}, {Spekkens}, {Wong}, {Lee-Waddell}, {Lagos},
		{Obreschkow}, {Ryan-Weber}, {Zwaan}, {Kilborn}, {Bekiaris}, {Bekki},
		{Bigiel}, {Boselli}, {Bosma}, {Catinella}, {Chauhan}, {Cluver}, {Colless},
		{Courtois}, {Crain}, {de Blok}, {D{\'e}nes}, {Duffy}, {Elagali}, {Fluke},
		{For}, {Heald}, {Henning}, {Hess}, {Holwerda}, {Howlett}, {Jarrett}, {Jones},
		{Jones}, {J{\'o}zsa}, {Jurek}, {J{\"u}tte}, {Kamphuis}, {Karachentsev},
		{Kerp}, {Kleiner}, {Kraan-Korteweg}, {L{\'o}pez-S{\'a}nchez}, {Madrid},
		{Meyer}, {Mould}, {Murugeshan}, {Norris}, {Oh}, {Oosterloo}, {Popping},
		{Putman}, {Reynolds}, {Rhee}, {Robotham}, {Ryder}, {Schr{\"o}der}, {Shao},
		{Stevens}, {Taylor}, {van{\^A} der Hulst}, {Verdes-Montenegro}, {Wakker},
		{Wang}, {Whiting}, {Winkel}, \& {Wolf}}]{Koribalski2020}
	{Koribalski}, B.~S., {Staveley-Smith}, L., {Westmeier}, T., {et~al.} 2020,
	\apss, 365, 118, \dodoi{10.1007/s10509-020-03831-4}
	
	\bibitem[{{Koss} {et~al.}(2021){Koss}, {Strittmatter}, {Lamperti}, {Shimizu},
		{Trakhtenbrot}, {Saintonge}, {Treister}, {Cicone}, {Mushotzky}, {Oh},
		{Ricci}, {Stern}, {Ananna}, {Bauer}, {Privon}, {B{\"a}r}, {De Breuck},
		{Harrison}, {Ichikawa}, {Powell}, {Rosario}, {Sanders}, {Schawinski}, {Shao},
		{Megan Urry}, \& {Veilleux}}]{Koss2021}
	{Koss}, M.~J., {Strittmatter}, B., {Lamperti}, I., {et~al.} 2021, \apjs, 252,
	29, \dodoi{10.3847/1538-4365/abcbfe}
	
	\bibitem[{{Krumholz}(2013)}]{Krumholz2013}
	{Krumholz}, M.~R. 2013, \mnras, 436, 2747, \dodoi{10.1093/mnras/stt1780}
	
	\bibitem[{{Krumholz} \& {Gnedin}(2011)}]{Krumholz2011}
	{Krumholz}, M.~R., \& {Gnedin}, N.~Y. 2011, \apj, 729, 36,
	\dodoi{10.1088/0004-637X/729/1/36}
	
	\bibitem[{{Leroy} {et~al.}(2008){Leroy}, {Walter}, {Brinks}, {Bigiel}, {de
			Blok}, {Madore}, \& {Thornley}}]{Leroy2008}
	{Leroy}, A.~K., {Walter}, F., {Brinks}, E., {et~al.} 2008, \aj, 136, 2782,
	\dodoi{10.1088/0004-6256/136/6/2782}
	
	\bibitem[{{Li} {et~al.}(2018){Li}, {Wang}, {Qian}, {Krco}, {Jiang}, {Yue},
		{Jin}, {Zhu}, {Pan}, {Nan}, \& {Dunning}}]{Li2018}
	{Li}, D., {Wang}, P., {Qian}, L., {et~al.} 2018, IEEE Microwave Magazine, 19,
	112, \dodoi{10.1109/MMM.2018.2802178}
	
	\bibitem[{{Lim} {et~al.}(2017){Lim}, {Mo}, {Lu}, {Wang}, \& {Yang}}]{Lim2017}
	{Lim}, S.~H., {Mo}, H.~J., {Lu}, Y., {Wang}, H., \& {Yang}, X. 2017, \mnras,
	470, 2982, \dodoi{10.1093/mnras/stx1462}
	
	\bibitem[{{Man} \& {Belli}(2018)}]{Man2018}
	{Man}, A., \& {Belli}, S. 2018, Nature Astronomy, 2, 695,
	\dodoi{10.1038/s41550-018-0558-1}
	
	\bibitem[{{Marinacci} {et~al.}(2018){Marinacci}, {Vogelsberger}, {Pakmor},
		{Torrey}, {Springel}, {Hernquist}, {Nelson}, {Weinberger}, {Pillepich},
		{Naiman}, \& {Genel}}]{Marinacci2018}
	{Marinacci}, F., {Vogelsberger}, M., {Pakmor}, R., {et~al.} 2018, \mnras, 480,
	5113, \dodoi{10.1093/mnras/sty2206}
	
	\bibitem[{{Martig} {et~al.}(2009){Martig}, {Bournaud}, {Teyssier}, \&
		{Dekel}}]{Martig2009}
	{Martig}, M., {Bournaud}, F., {Teyssier}, R., \& {Dekel}, A. 2009, \apj, 707,
	250, \dodoi{10.1088/0004-637X/707/1/250}
	
	\bibitem[{{Martin} {et~al.}(2010){Martin}, {Papastergis}, {Giovanelli},
		{Haynes}, {Springob}, \& {Stierwalt}}]{Martin2010}
	{Martin}, A.~M., {Papastergis}, E., {Giovanelli}, R., {et~al.} 2010, \apj, 723,
	1359, \dodoi{10.1088/0004-637X/723/2/1359}
	
	\bibitem[{{McCarthy} {et~al.}(2011){McCarthy}, {Schaye}, {Bower}, {Ponman},
		{Booth}, {Dalla Vecchia}, \& {Springel}}]{McCarthy2011}
	{McCarthy}, I.~G., {Schaye}, J., {Bower}, R.~G., {et~al.} 2011, \mnras, 412,
	1965, \dodoi{10.1111/j.1365-2966.2010.18033.x}
	
	\bibitem[{{Naiman} {et~al.}(2018){Naiman}, {Pillepich}, {Springel},
		{Ramirez-Ruiz}, {Torrey}, {Vogelsberger}, {Pakmor}, {Nelson}, {Marinacci},
		{Hernquist}, {Weinberger}, \& {Genel}}]{Naiman2018}
	{Naiman}, J.~P., {Pillepich}, A., {Springel}, V., {et~al.} 2018, \mnras, 477,
	1206, \dodoi{10.1093/mnras/sty618}
	
	\bibitem[{{Nelson} {et~al.}(2015){Nelson}, {Pillepich}, {Genel},
		{Vogelsberger}, {Springel}, {Torrey}, {Rodriguez-Gomez}, {Sijacki}, {Snyder},
		{Griffen}, {Marinacci}, {Blecha}, {Sales}, {Xu}, \& {Hernquist}}]{Nelson2015}
	{Nelson}, D., {Pillepich}, A., {Genel}, S., {et~al.} 2015, Astronomy and
	Computing, 13, 12, \dodoi{10.1016/j.ascom.2015.09.003}
	
	\bibitem[{{Nelson} {et~al.}(2018){Nelson}, {Pillepich}, {Springel},
		{Weinberger}, {Hernquist}, {Pakmor}, {Genel}, {Torrey}, {Vogelsberger},
		{Kauffmann}, {Marinacci}, \& {Naiman}}]{Nelson2018}
	{Nelson}, D., {Pillepich}, A., {Springel}, V., {et~al.} 2018, \mnras, 475, 624,
	\dodoi{10.1093/mnras/stx3040}
	
	\bibitem[{{Nelson} {et~al.}(2019{\natexlab{a}}){Nelson}, {Springel},
		{Pillepich}, {Rodriguez-Gomez}, {Torrey}, {Genel}, {Vogelsberger}, {Pakmor},
		{Marinacci}, {Weinberger}, {Kelley}, {Lovell}, {Diemer}, \&
		{Hernquist}}]{Nelson2019}
	{Nelson}, D., {Springel}, V., {Pillepich}, A., {et~al.} 2019{\natexlab{a}},
	Computational Astrophysics and Cosmology, 6, 2,
	\dodoi{10.1186/s40668-019-0028-x}
	
	\bibitem[{{Nelson} {et~al.}(2019{\natexlab{b}}){Nelson}, {Pillepich},
		{Springel}, {Pakmor}, {Weinberger}, {Genel}, {Torrey}, {Vogelsberger},
		{Marinacci}, \& {Hernquist}}]{Nelson2019b}
	{Nelson}, D., {Pillepich}, A., {Springel}, V., {et~al.} 2019{\natexlab{b}},
	\mnras, 490, 3234, \dodoi{10.1093/mnras/stz2306}
	
	\bibitem[{{Pakmor} {et~al.}(2016){Pakmor}, {Springel}, {Bauer}, {Mocz},
		{Munoz}, {Ohlmann}, {Schaal}, \& {Zhu}}]{Pakmor2016}
	{Pakmor}, R., {Springel}, V., {Bauer}, A., {et~al.} 2016, \mnras, 455, 1134,
	\dodoi{10.1093/mnras/stv2380}
	
	\bibitem[{{Pillepich} {et~al.}(2018){Pillepich}, {Springel}, {Nelson}, {Genel},
		{Naiman}, {Pakmor}, {Hernquist}, {Torrey}, {Vogelsberger}, {Weinberger}, \&
		{Marinacci}}]{Pillepich2018}
	{Pillepich}, A., {Springel}, V., {Nelson}, D., {et~al.} 2018, \mnras, 473,
	4077, \dodoi{10.1093/mnras/stx2656}
	
	\bibitem[{{Pillepich} {et~al.}(2019){Pillepich}, {Nelson}, {Springel},
		{Pakmor}, {Torrey}, {Weinberger}, {Vogelsberger}, {Marinacci}, {Genel}, {van
			der Wel}, \& {Hernquist}}]{Pillepich2019}
	{Pillepich}, A., {Nelson}, D., {Springel}, V., {et~al.} 2019, \mnras, 490,
	3196, \dodoi{10.1093/mnras/stz2338}
	
	\bibitem[{{Piotrowska} {et~al.}(2022){Piotrowska}, {Bluck}, {Maiolino}, \&
		{Peng}}]{Piotrowska2022}
	{Piotrowska}, J.~M., {Bluck}, A. F.~L., {Maiolino}, R., \& {Peng}, Y. 2022,
	\mnras, 512, 1052, \dodoi{10.1093/mnras/stab3673}
	
	\bibitem[{{Rahmati} {et~al.}(2013){Rahmati}, {Pawlik}, {Rai{\v{c}}evi{\'c}}, \&
		{Schaye}}]{Rahmati2013}
	{Rahmati}, A., {Pawlik}, A.~H., {Rai{\v{c}}evi{\'c}}, M., \& {Schaye}, J. 2013,
	\mnras, 430, 2427, \dodoi{10.1093/mnras/stt066}
	
	\bibitem[{{Saintonge} \& {Catinella}(2022)}]{Saintonge2022}
	{Saintonge}, A., \& {Catinella}, B. 2022, \araa, 60, 319,
	\dodoi{10.1146/annurev-astro-021022-043545}
	
	\bibitem[{{Saintonge} {et~al.}(2016){Saintonge}, {Catinella}, {Cortese},
		{Genzel}, {Giovanelli}, {Haynes}, {Janowiecki}, {Kramer}, {Lutz},
		{Schiminovich}, {Tacconi}, {Wuyts}, \& {Accurso}}]{Saintonge2016}
	{Saintonge}, A., {Catinella}, B., {Cortese}, L., {et~al.} 2016, \mnras, 462,
	1749, \dodoi{10.1093/mnras/stw1715}
	
	\bibitem[{{Saintonge} {et~al.}(2017){Saintonge}, {Catinella}, {Tacconi},
		{Kauffmann}, {Genzel}, {Cortese}, {Dav{\'e}}, {Fletcher},
		{Graci{\'a}-Carpio}, {Kramer}, {Heckman}, {Janowiecki}, {Lutz}, {Rosario},
		{Schiminovich}, {Schuster}, {Wang}, {Wuyts}, {Borthakur}, {Lamperti}, \&
		{Roberts-Borsani}}]{Saintonge2017}
	{Saintonge}, A., {Catinella}, B., {Tacconi}, L.~J., {et~al.} 2017, \apjs, 233,
	22, \dodoi{10.3847/1538-4365/aa97e0}
	
	\bibitem[{{Salim} {et~al.}(2018){Salim}, {Boquien}, \& {Lee}}]{Salim2018}
	{Salim}, S., {Boquien}, M., \& {Lee}, J.~C. 2018, \apj, 859, 11,
	\dodoi{10.3847/1538-4357/aabf3c}
	
	\bibitem[{{Schaye} {et~al.}(2015){Schaye}, {Crain}, {Bower}, {Furlong},
		{Schaller}, {Theuns}, {Dalla Vecchia}, {Frenk}, {McCarthy}, {Helly},
		{Jenkins}, {Rosas-Guevara}, {White}, {Baes}, {Booth}, {Camps}, {Navarro},
		{Qu}, {Rahmati}, {Sawala}, {Thomas}, \& {Trayford}}]{Schaye2015}
	{Schaye}, J., {Crain}, R.~A., {Bower}, R.~G., {et~al.} 2015, \mnras, 446, 521,
	\dodoi{10.1093/mnras/stu2058}
	
	\bibitem[{{Schmidt}(1959)}]{Schmidt1959}
	{Schmidt}, M. 1959, \apj, 129, 243, \dodoi{10.1086/146614}
	
	\bibitem[{{Shangguan} \& {Ho}(2019)}]{Shangguan2019}
	{Shangguan}, J., \& {Ho}, L.~C. 2019, \apj, 873, 90,
	\dodoi{10.3847/1538-4357/ab0555}
	
	\bibitem[{{Shangguan} {et~al.}(2020){Shangguan}, {Ho}, {Bauer}, {Wang}, \&
		{Treister}}]{Shangguan2020}
	{Shangguan}, J., {Ho}, L.~C., {Bauer}, F.~E., {Wang}, R., \& {Treister}, E.
	2020, \apj, 899, 112, \dodoi{10.3847/1538-4357/aba8a1}
	
	\bibitem[{{Shangguan} {et~al.}(2018){Shangguan}, {Ho}, \&
		{Xie}}]{Shangguan2018}
	{Shangguan}, J., {Ho}, L.~C., \& {Xie}, Y. 2018, \apj, 854, 158,
	\dodoi{10.3847/1538-4357/aaa9be}
	
	\bibitem[{{Sijacki} {et~al.}(2007){Sijacki}, {Springel}, {Di Matteo}, \&
		{Hernquist}}]{Sijacki2007}
	{Sijacki}, D., {Springel}, V., {Di Matteo}, T., \& {Hernquist}, L. 2007,
	\mnras, 380, 877, \dodoi{10.1111/j.1365-2966.2007.12153.x}
	
	\bibitem[{{Sijacki} {et~al.}(2015){Sijacki}, {Vogelsberger}, {Genel},
		{Springel}, {Torrey}, {Snyder}, {Nelson}, \& {Hernquist}}]{Sijacki2015}
	{Sijacki}, D., {Vogelsberger}, M., {Genel}, S., {et~al.} 2015, \mnras, 452,
	575, \dodoi{10.1093/mnras/stv1340}
	
	\bibitem[{{Springel}(2010)}]{Springel2010}
	{Springel}, V. 2010, \mnras, 401, 791, \dodoi{10.1111/j.1365-2966.2009.15715.x}
	
	\bibitem[{{Springel} \& {Hernquist}(2003)}]{Springel2003}
	{Springel}, V., \& {Hernquist}, L. 2003, \mnras, 339, 289,
	\dodoi{10.1046/j.1365-8711.2003.06206.x}
	
	\bibitem[{{Springel} {et~al.}(2018){Springel}, {Pakmor}, {Pillepich},
		{Weinberger}, {Nelson}, {Hernquist}, {Vogelsberger}, {Genel}, {Torrey},
		{Marinacci}, \& {Naiman}}]{Springel2018}
	{Springel}, V., {Pakmor}, R., {Pillepich}, A., {et~al.} 2018, \mnras, 475, 676,
	\dodoi{10.1093/mnras/stx3304}
	
	\bibitem[{{Sternberg} {et~al.}(2014){Sternberg}, {Le Petit}, {Roueff}, \& {Le
			Bourlot}}]{Sternberg2014}
	{Sternberg}, A., {Le Petit}, F., {Roueff}, E., \& {Le Bourlot}, J. 2014, \apj,
	790, 10, \dodoi{10.1088/0004-637X/790/1/10}
	
	\bibitem[{{Terrazas} {et~al.}(2020){Terrazas}, {Bell}, {Pillepich}, {Nelson},
		{Somerville}, {Genel}, {Weinberger}, {Habouzit}, {Li}, {Hernquist}, \&
		{Vogelsberger}}]{Terrazas2020}
	{Terrazas}, B.~A., {Bell}, E.~F., {Pillepich}, A., {et~al.} 2020, \mnras, 493,
	1888, \dodoi{10.1093/mnras/staa374}
	
	\bibitem[{{Thomas} {et~al.}(2019){Thomas}, {Dav{\'e}},
		{Angl{\'e}s-Alc{\'a}zar}, \& {Jarvis}}]{Thomas2019}
	{Thomas}, N., {Dav{\'e}}, R., {Angl{\'e}s-Alc{\'a}zar}, D., \& {Jarvis}, M.
	2019, \mnras, 487, 5764, \dodoi{10.1093/mnras/stz1703}
	
	\bibitem[{{Varma} {et~al.}(2022){Varma}, {Huertas-Company}, {Pillepich},
		{Nelson}, {Rodriguez-Gomez}, {Dekel}, {Faber}, {Iglesias-Navarro}, {Koo}, \&
		{Primack}}]{Varma2022}
	{Varma}, S., {Huertas-Company}, M., {Pillepich}, A., {et~al.} 2022, \mnras,
	509, 2654, \dodoi{10.1093/mnras/stab3149}
	
	\bibitem[{{Verheijen} {et~al.}(2008){Verheijen}, {Oosterloo}, {van Cappellen},
		{Bakker}, {Ivashina}, \& {van der Hulst}}]{Verheijen2008}
	{Verheijen}, M.~A.~W., {Oosterloo}, T.~A., {van Cappellen}, W.~A., {et~al.}
	2008, in American Institute of Physics Conference Series, Vol. 1035, The
	Evolution of Galaxies Through the Neutral Hydrogen Window, ed. R.~{Minchin}
	\& E.~{Momjian}, 265--271, \dodoi{10.1063/1.2973599}
	
	\bibitem[{{Vito} {et~al.}(2014){Vito}, {Maiolino}, {Santini}, {Brusa},
		{Comastri}, {Cresci}, {Farrah}, {Franceschini}, {Gilli}, {Granato},
		{Gruppioni}, {Lutz}, {Mannucci}, {Pozzi}, {Rosario}, {Scott}, {Viero}, \&
		{Vignali}}]{Vito2014}
	{Vito}, F., {Maiolino}, R., {Santini}, P., {et~al.} 2014, \mnras, 441, 1059,
	\dodoi{10.1093/mnras/stu637}
	
	\bibitem[{{Vogelsberger} {et~al.}(2014){Vogelsberger}, {Genel}, {Springel},
		{Torrey}, {Sijacki}, {Xu}, {Snyder}, {Nelson}, \&
		{Hernquist}}]{Vogelsberger2014}
	{Vogelsberger}, M., {Genel}, S., {Springel}, V., {et~al.} 2014, \mnras, 444,
	1518, \dodoi{10.1093/mnras/stu1536}
	
	\bibitem[{{Wang} {et~al.}(2020){Wang}, {Catinella}, {Saintonge}, {Pan},
		{Serra}, \& {Shao}}]{Wang2020}
	{Wang}, J., {Catinella}, B., {Saintonge}, A., {et~al.} 2020, \apj, 890, 63,
	\dodoi{10.3847/1538-4357/ab68dd}
	
	\bibitem[{{Wang} {et~al.}(2021){Wang}, {Staveley-Smith}, {Westmeier},
		{Catinella}, {Shao}, {Reynolds}, {For}, {Lee}, {Liang}, {Wang}, {Elagali},
		{D{\'e}nes}, {Kleiner}, {Koribalski}, {Lee-Waddell}, {Oh}, {Rhee}, {Serra},
		{Spekkens}, {Wong}, {Bekki}, {Bigiel}, {Courtois}, {Hess}, {Holwerda},
		{McQuinn}, {Pandey-Pommier}, {van der Hulst}, \&
		{Verdes-Montenegro}}]{Wang2021}
	{Wang}, J., {Staveley-Smith}, L., {Westmeier}, T., {et~al.} 2021, \apj, 915,
	70, \dodoi{10.3847/1538-4357/abfc52}
	
	\bibitem[{{Wang} {et~al.}(2022){Wang}, {Wang}, {For}, {Lee}, {Reynolds}, {Lin},
		{Staveley-Smith}, {Shao}, {Wong}, {Catinella}, {Serra}, {Verdes-Montenegro},
		{Westmeier}, {Lee-Waddell}, {Koribalski}, {Murugeshan}, {Elagali}, {Kleiner},
		{Rhee}, {Bigiel}, {Bosma}, {Holwerda}, {Oh}, \& {Spekkens}}]{Wang2022}
	{Wang}, S., {Wang}, J., {For}, B.-Q., {et~al.} 2022, \apj, 927, 66,
	\dodoi{10.3847/1538-4357/ac4270}
	
	\bibitem[{{Ward} {et~al.}(2022){Ward}, {Harrison}, {Costa}, \&
		{Mainieri}}]{Ward2022}
	{Ward}, S.~R., {Harrison}, C.~M., {Costa}, T., \& {Mainieri}, V. 2022, \mnras,
	514, 2936, \dodoi{10.1093/mnras/stac1219}
	
	\bibitem[{{Weinberger} {et~al.}(2017){Weinberger}, {Springel}, {Hernquist},
		{Pillepich}, {Marinacci}, {Pakmor}, {Nelson}, {Genel}, {Vogelsberger},
		{Naiman}, \& {Torrey}}]{Weinberger2017}
	{Weinberger}, R., {Springel}, V., {Hernquist}, L., {et~al.} 2017, \mnras, 465,
	3291, \dodoi{10.1093/mnras/stw2944}
	
	\bibitem[{{Weinberger} {et~al.}(2018){Weinberger}, {Springel}, {Pakmor},
		{Nelson}, {Genel}, {Pillepich}, {Vogelsberger}, {Marinacci}, {Naiman},
		{Torrey}, \& {Hernquist}}]{Weinberger2018}
	{Weinberger}, R., {Springel}, V., {Pakmor}, R., {et~al.} 2018, \mnras, 479,
	4056, \dodoi{10.1093/mnras/sty1733}
	
	\bibitem[{{Wong} \& {Blitz}(2002)}]{Wong2002}
	{Wong}, T., \& {Blitz}, L. 2002, \apj, 569, 157, \dodoi{10.1086/339287}
	
	\bibitem[{{York} {et~al.}(2000){York}, {Adelman}, {Anderson}, {Anderson},
		{Annis}, {Bahcall}, {Bakken}, {Barkhouser}, {Bastian}, {Berman}, {Boroski},
		{Bracker}, {Briegel}, {Briggs}, {Brinkmann}, {Brunner}, {Burles}, {Carey},
		{Carr}, {Castander}, {Chen}, {Colestock}, {Connolly}, {Crocker}, {Csabai},
		{Czarapata}, {Davis}, {Doi}, {Dombeck}, {Eisenstein}, {Ellman}, {Elms},
		{Evans}, {Fan}, {Federwitz}, {Fiscelli}, {Friedman}, {Frieman}, {Fukugita},
		{Gillespie}, {Gunn}, {Gurbani}, {de Haas}, {Haldeman}, {Harris}, {Hayes},
		{Heckman}, {Hennessy}, {Hindsley}, {Holm}, {Holmgren}, {Huang}, {Hull},
		{Husby}, {Ichikawa}, {Ichikawa}, {Ivezi{\'c}}, {Kent}, {Kim}, {Kinney},
		{Klaene}, {Kleinman}, {Kleinman}, {Knapp}, {Korienek}, {Kron}, {Kunszt},
		{Lamb}, {Lee}, {Leger}, {Limmongkol}, {Lindenmeyer}, {Long}, {Loomis},
		{Loveday}, {Lucinio}, {Lupton}, {MacKinnon}, {Mannery}, {Mantsch}, {Margon},
		{McGehee}, {McKay}, {Meiksin}, {Merelli}, {Monet}, {Munn}, {Narayanan},
		{Nash}, {Neilsen}, {Neswold}, {Newberg}, {Nichol}, {Nicinski}, {Nonino},
		{Okada}, {Okamura}, {Ostriker}, {Owen}, {Pauls}, {Peoples}, {Peterson},
		{Petravick}, {Pier}, {Pope}, {Pordes}, {Prosapio}, {Rechenmacher}, {Quinn},
		{Richards}, {Richmond}, {Rivetta}, {Rockosi}, {Ruthmansdorfer}, {Sandford},
		{Schlegel}, {Schneider}, {Sekiguchi}, {Sergey}, {Shimasaku}, {Siegmund},
		{Smee}, {Smith}, {Snedden}, {Stone}, {Stoughton}, {Strauss}, {Stubbs},
		{SubbaRao}, {Szalay}, {Szapudi}, {Szokoly}, {Thakar}, {Tremonti}, {Tucker},
		{Uomoto}, {Vanden Berk}, {Vogeley}, {Waddell}, {Wang}, {Watanabe},
		{Weinberg}, {Yanny}, {Yasuda}, \& {SDSS Collaboration}}]{York2000}
	{York}, D.~G., {Adelman}, J., {Anderson}, John~E., J., {et~al.} 2000, \aj, 120,
	1579, \dodoi{10.1086/301513}
	
	\bibitem[{{Zinger} {et~al.}(2020){Zinger}, {Pillepich}, {Nelson}, {Weinberger},
		{Pakmor}, {Springel}, {Hernquist}, {Marinacci}, \&
		{Vogelsberger}}]{Zinger2020}
	{Zinger}, E., {Pillepich}, A., {Nelson}, D., {et~al.} 2020, \mnras, 499, 768,
	\dodoi{10.1093/mnras/staa2607}
	
	\bibitem[{{Zolotov} {et~al.}(2015){Zolotov}, {Dekel}, {Mandelker}, {Tweed},
		{Inoue}, {DeGraf}, {Ceverino}, {Primack}, {Barro}, \& {Faber}}]{Zolotov2015}
	{Zolotov}, A., {Dekel}, A., {Mandelker}, N., {et~al.} 2015, \mnras, 450, 2327,
	\dodoi{10.1093/mnras/stv740}
	
	\bibitem[{{Zwaan} {et~al.}(2005){Zwaan}, {Meyer}, {Staveley-Smith}, \&
		{Webster}}]{Zwaan2005}
	{Zwaan}, M.~A., {Meyer}, M.~J., {Staveley-Smith}, L., \& {Webster}, R.~L. 2005,
	\mnras, 359, L30, \dodoi{10.1111/j.1745-3933.2005.00029.x}
	
\end{thebibliography}

\end{document}